\documentclass[12pt]{article}

\usepackage{amsmath}

\usepackage{amssymb}
\usepackage{bm}
\usepackage{braket}

\usepackage[pdftex]{graphicx}

\usepackage{here}
\usepackage{comment}

\begin{document}

\baselineskip=17pt

\begin{titlepage}
\rightline{\tt arXiv:2305.13103}
\begin{center}
\vskip 1.5cm
\baselineskip=22pt
{\Large \bf {Correlation functions involving Dirac fields}}\\
{\Large \bf {from homotopy algebras II: the interacting theory}}
\end{center}
\begin{center}
\vskip 1.0cm
{\large Keisuke Konosu}
\vskip 1.0cm
{\it {Graduate School of Arts and Sciences, The University of Tokyo}}\\
{\it {3-8-1 Komaba, Meguro-ku, Tokyo 153-8902, Japan}}\\
konosu-keisuke@g.ecc.u-tokyo.ac.jp
\vskip 2.0cm

{\bf Abstract}
\end{center}

\noindent
We extend the formula for correlation functions of free scalar field theories and Dirac field theories
in terms of quantum $A_{\infty}$ algebras presented in arXiv:2305.11634
to general scalar-Dirac systems. We obtain the result that the same formula as in the previous paper holds in this case. We show that correlation functions from our formula satisfy the Schwinger-Dyson equations. We therefore confirm that correlation functions from our formula express correlation functions from the ordinary approach of quantum field theory. 

\end{titlepage}

\tableofcontents

\section{Introduction}
\setcounter{equation}{0}
Homotopy algebras such as $A_{\infty}$ algebras~\cite{Stasheff:I, Stasheff:II, Getzler-Jones, Markl, Penkava:1994mu, Gaberdiel:1997ia} and $L_{\infty}$ algebras~\cite{Zwiebach:1992ie, Markl:1997bj} are powerful tools  to construct  actions of string field theory. We can also integrate out fields~\cite{Sen:2016qap, Erbin:2020eyc, Koyama:2020qfb, Arvanitakis:2020rrk, Arvanitakis:2021ecw}, calculate scattering amplitudes\footnote{The tree-level S-matrix of superstring field theory is discussed, for example, \cite{Konopka:2015tta} and \cite{Kunitomo:2020xrl} in this context.}~\cite{Kajiura:2003ax}, and relate between covariant and light-cone string field theories~\cite{Erler:2020beb}. These applications are very important contributions to string field theory. 

Homotopy algebras can be applied not only for string field theory
but also for quantum field theory~\cite{Hohm:2017pnh, Jurco:2018sby, Nutzi:2018vkl, Arvanitakis:2019ald, Macrelli:2019afx, Jurco:2019yfd, Saemann:2020oyz}.  Expressions of Feynman diagrams can be reproduced algebraically
in the framework of homotopy algebras~\cite{Kajiura:2003ax, Doubek:2017naz, Masuda:2020tfa}, and their descriptions are known to be the same as in the case of string field theory. We expect that establishing the full description of quantum field theory in homotopy algebras will lead to the full description of string field theory by the universal descriptions of homotopy algebras. 

In the previous paper~\cite{Konosu/Okawa_2022}, we extend the scalar correlation function formula\footnote{Related works are seen in \cite{Gwilliam:2012jg} and \cite{Chiaffrino:2021pob}. In these references, 
 correlation functions in the framework of the Batalin-Vilkovisky formalism~\cite{Batalin:1981jr, Batalin:1983ggl, Schwarz:1992nx} are discussed. See also recent development~\cite{DimitrijevicCiric:2023hua}. }
 proposed in \cite{Okawa:2022sjf} to the free Dirac fields by introducing string-field-theory-like convention.\footnote{
Related discussions can be found in~subsection~3.3 of~\cite{Kajiura:2003ax}
and appendix~A of~\cite{Masuda:2020tfa}.} In this paper, we generalize this result to interacting theories that include both scalar and Dirac fields. As an example, we construct the modified Yukawa theory, calculate the one-loop correction of the one-point functions and the two-point functions, and confirm that these reproduce the ordinary Feynman diagram calculations using our new formulas. Then, we show that correlation functions from our formula satisfy  Schwinger-Dyson equations.

The rest of the paper is organized as follows. In section~\ref{review-section}, we briefly explain notations and review the previous results in~\cite{Konosu/Okawa_2022}. In section~\ref{mod-Yukawa}, we calculate the one-loop correction of the one-point functions and the two-point functions and see that our formula reproduces the correct correlation functions of the modified Yukawa theory. In section~\ref{Schwinger-Dyson}, we show that correlation functions from our formula satisfy the Schwinger-Dyson equations. Section~\ref{conclusion-section} is devoted to conclusions and discussion. In Appendix~\ref{Yukawa}, we calculate the tree-level scattering amplitudes of the Yukawa theory in our string-field-theory-like formulation.

\section{Notations and reviews} \label{review-section}
Following \cite{Konosu/Okawa_2022}, we state notations and review the correlation functions of  scalar field theories and free Dirac theories in $d$ dimensions in terms of $A_{\infty}$ algebras. Our notations on quantum field theory are mostly based on \cite{Srednicki:2007qs}. 
\subsection{Notations on $A_{\infty}$ algebras}
\setcounter{equation}{0}
In this subsection, we briefly state our notations on $A_{\infty}$ algebras.
We consider the vector space
\begin{equation}
  \mathcal{H} = \bigoplus_{n\in\mathbb{Z}} \mathcal{H}_{n} \,. \label{graded-vector-space}
\end{equation}
The classical action is written in terms of degree-even elements of $\mathcal{H}_1$. We denote it by $\Phi$ and  consider an action of the form\footnote{In general, we use a master action in the Batalin-Vilkovisky formalism. In this paper, we only consider theories without gauge symmetry, whose actions coincide with the classical actions.}
\begin{equation}
S = {}-\frac{1}{2} \, \omega \, ( \, \Phi, Q \, \Phi \, )
-\sum_{n=0}^\infty \, \frac{1}{n+1} \,
\omega \, ( \, \Phi \,, m_n \, ( \, \Phi \otimes \ldots \otimes \Phi \, ) \, ) \,.
\end{equation}
The symplectic form $\omega \, ( \, \Phi_1 \,, \Phi_2 \, )$  is
defined to satisfy the following property:
\begin{equation}
  \omega \, ( \, \Phi_1 \,, \Phi_2 \, )
  = {}-(-1)^{\mathrm{deg} (\Phi_1) \, \mathrm{deg} (\Phi_2)} \,
  \omega \, ( \,\Phi_2 \,, \Phi_1 \, )\,,
\label{omega-symplectic}
\end{equation}
where $\Phi_1$ and $\Phi_2$ are elements of $\mathcal{H}$.
Here, we introduced $\mathbb{Z}_{2}$ grading and denote the degree of $\Phi$ by $\mathrm{deg} \, (\Phi)$:
\begin{equation}
	\mathrm{deg}(\Phi)=\left\{
	\begin{array}{ll}
	0 & (\Phi\,\,\mathrm{:\,degree\,even}\,) \hspace{0.6 cm} (\mathrm{mod}\,2)\\
	1 & (\Phi\,\,\mathrm{:\,degree\,odd}\,) \hspace{0.7 cm}\, (\mathrm{mod}\,2)\,.
	\end{array}
	\right.
\end{equation}
The operator $Q$ is  degree odd from $\mathcal{H}$ to $\mathcal{H}$ and describes the kinetic terms of the free theory,
and operators $m_{n}$ describe interactions which involves $n+1$ fields. Note that operators $m_{n}$
is a degree-odd map from $\mathcal{H}^{\otimes n}$ to $\mathcal{H}$,
where
\begin{equation}
  \mathcal{H}^{\otimes n}
  = \underbrace{\, \mathcal{H}\otimes\mathcal{H}\otimes\ldots\otimes\mathcal{H} \,}_{n}\,,
\end{equation}
for $n > 0$. The space $\mathcal{H}^{\otimes 0}$ is defined to be a one-dimensional vector space
constructed with a single basis vector $\bf{1}$ multiplied by complex numbers. This basis vector is degree even and serves as the unit of tensor products as follows:\begin{equation}
{\bf 1} \otimes \Phi = \Phi \,, \quad \Phi \otimes {\bf 1} = \Phi\,,
\end{equation}
where $\Phi$ is an element of $\mathcal{H}$. We assume the cyclicity given by 
\begin{equation}
\omega \, ( \, \Phi_1 \,, Q \,\Phi_2 \, )
= {}-(-1)^{\mathrm{deg}(\Phi_1)} \,
\omega \, ( \, Q \Phi_1 \,, \Phi_{2} \, ) 
\end{equation}
and
\begin{equation}
\omega \, ( \, \Phi_1 \,, m_n \, ( \, \Phi_2 \otimes \ldots \otimes \Phi_{n+1} \, ) \, )
= {}-(-1)^{\mathrm{deg}(\Phi_1)} \,
\omega \, ( \, m_n ( \, \Phi_1 \otimes \ldots \otimes \Phi_n \, ) \,, \Phi_{n+1} \, ) \,,
\label{cyclic}
\end{equation}
where $\Phi_1, \ldots, \Phi_n,$ and $\Phi_{n+1}$ are elements of $\mathcal{H}$.

Let us next introduce the coalgebra representation.
We define
\begin{equation}
  T\mathcal{H} = \mathcal{H}^{\otimes 0} \oplus \mathcal{H} \oplus \mathcal{H}^{\otimes 2} \oplus \mathcal{H}^{\otimes 3}\oplus\ldots\,,
\end{equation}
and consider linear operators acting on this space.
First, we consider $c_n$ which is a map from $\mathcal{H}^{\otimes n}$ to $\mathcal{H}$.
Associated with this operator, coderivation $\bm{c}_n$ acting on $T \mathcal{H}$ is defined by
\begin{align}
\bm{c}_n \pi_m & = 0 \quad \text{for} \quad m < n \,, \\
\bm{c}_n \pi_n & = c_n \, \pi_n \,, \\
\bm{c}_n \pi_{n+1} & = ( c_n \otimes \mathbb{I} + \mathbb{I} \otimes c_n ) \, \pi_{n+1} \,, \\
\bm{c}_n \pi_m
& = ( c_n \otimes \mathbb{I}^{\otimes(m-n)}
+\sum_{k=1}^{m-n-1} \mathbb{I}^{\otimes k} \otimes c_n \otimes \mathbb{I}^{\otimes (m-n-k)}
+\mathbb{I}^{\otimes(m-n)} \otimes c_n ) \, \pi_m \quad \text{for} \quad m > n+1 \,,
\end{align}
where $\pi_n$ is the projection operator from $T\mathcal{H}$ onto $\mathcal{H}^{\otimes n}$,
$\mathbb{I}$ is the identity operator on $\mathcal{H}$,
and we abbriviate 
\begin{equation}
 \underbrace{\, \mathbb{I} \otimes \mathbb{I} \otimes \ldots \otimes \mathbb{I} \,}_n \,.
\end{equation}
by $\mathbb{I}^{\otimes n}$.
We can compactly express the $A_\infty$ relations as
\begin{equation}
	\left({\bf Q}+\sum_{n=0}^{\infty}{\bm m}_{n}\right)^2 = 0 \,,
\end{equation}
where ${\bf Q}$ and ${\bm m}_{n}$ are coderivations associated with the operators $Q$ and $m_{n}$, respectively. 

When we discuss quantum field theory in terms of $A_{\infty}$ algebras, the  projection operator $P$ onto a subspace of $\mathcal{H}$ is important.
We define $P$ to be degree even. In the coalgebra representation, the projection operator ${\bf P}$ acting on $T \mathcal{H}$ associated with $P$ is defined by
\begin{equation}
\begin{split}
{\bf P} \, \pi_0 & = \pi_0 \,, \\
{\bf P} \, \pi_n & =  (\,\underbrace{\, P \otimes P \otimes \ldots \otimes P \,}_n\,) \, \pi_n
\end{split}
\end{equation}
for $n > 0$.
Furthermore, we also need the \textit{contracting homotopy} $h$, which is defined to satisfy 
\begin{equation}
Q\, h+h \,Q=\mathbb{I}-P,  \quad h \,P=0, \quad P \,h=0, \quad h^2=0\,.\label{HK}
\end{equation}
In the coalgebra representations, the contracting homotopy $h$ is defined as follows: 
\begin{align}
\bm{h} \, \pi_0 & = 0 \,, \\
\bm{h} \, \pi_1 & = h \, \pi_1 \,, \\
\bm{h} \, \pi_2 & = ( \, h \otimes P +\mathbb{I} \otimes h \, ) \, \pi_2 \,, \\
\bm{h} \, \pi_n & = \Bigl( h \otimes P^{\otimes(n-1)}
+\sum_{m=1}^{n-2} \mathbb{I}^{\otimes m} \otimes h \otimes P^{\otimes(n-m-1)}
+\mathbb{I}^{\otimes(n-1)} \otimes h \Bigr) \, \pi_n\qquad (n > 2)\,.
\end{align}

\subsection{Scalar fields}\label{scalar-section}
In this subsection, we consider scalar field theory.
In general, we can decompose the vector space $\mathcal{H}$ as 
\begin{equation}
	\mathcal{H}=\bigoplus_{n\in\mathbb{Z}}\mathcal{H}_{n}
\end{equation} 
as explained in the previous subsection.
When we consider theories without gauge symmetry, we only need two sectors which are denoted by $\mathcal{H}_{1}$ and $\mathcal{H}_{2}$:
\begin{equation}
	\mathcal{H}=\mathcal{H}_{1}\oplus\mathcal{H}_{2}\,.
\end{equation}
The element $\Phi$ of $\mathcal{H}_{1}$ can be expanded as
\begin{equation}
\Phi = \int d^d x \, \varphi (x) \, c(x) \,, \label{scalar-expansion}
\end{equation}
where $\varphi (x)$ is a real scalar field  and $c(x)$ is the basis vector of $\mathcal{H}_{1}$. We define $\varphi (x)$ and $c(x)$ as degree even.
The symplectic form $\omega$ is defined by
\begin{equation}
\biggl(
\begin{array}{cc}
\omega \, ( \, c(x_1) \,, c(x_2) \, ) & \omega \, ( \, c(x_1) \,, d(x_2) \, ) \\
\omega \, ( \, d(x_1) \,, c(x_2) \, ) & \omega \, ( \, d(x_1) \,, d(x_2) \, )
\end{array}
\biggr)
= \biggl(
\begin{array}{cc}
0 & \delta^d ( x_1-x_2 ) \\
{}-\delta^d ( x_1-x_2 ) & 0
\end{array}
\biggr) \,,
\end{equation}
where we introduced the degree-odd basis vector $d(x)$ for $\mathcal{H}_{2}$.  We then define the operator $Q$ by
\begin{equation}
Q \, c(x) = ( \, {}-\partial^2 +M^2 \, ) \, d(x) \,, \qquad
Q \, d(x) = 0 \,,
\end{equation}
where $M$ is the mass of the scalar field to obtain
\begin{equation}
	\begin{split}
 		 S &= {}-\frac{1}{2} \, \omega \, ( \, \Phi, Q \, \Phi \, )\\
		 &={}-\frac{1}{2} \int d^d x \, \bigl[ \, \partial_\mu \varphi (x) \, \partial^\mu \varphi (x) +M^2 \, \varphi (x)^2 \, \bigr]
  	\end{split}
\end{equation}
with $\Phi$ in~\eqref{scalar-expansion}. Note that the cyclic property of $Q$ is satisfied:
\begin{equation}
\omega \, ( \, Q \, \Phi_1 \,, \Phi_2 \, )
= {}-(-1)^{\mathrm{deg} (\Phi_1)} \, \omega \, ( \, \Phi_1 \,, Q \, \Phi_2 \, )\,.
\end{equation}

Let us consider correlation functions in terms of homotopy algebras. According to \cite{Okawa:2022sjf}, we set
\begin{equation}
	P=0\,.
\end{equation}
Then, the definitions for the contracting homotopy~\eqref{HK} becomes
\begin{equation}
	Q \, h +h \, Q = \mathbb{I} \,, \qquad
	h^2 = 0 \,.
\end{equation}
We define
\begin{equation}
h \, c(x) = 0 \,, \qquad
h \, d(x) = \int d^d y \, \Delta (x-y) \, c (y) \,,
\label{scalar-h}
\end{equation}
where $\Delta(x-y)$ is the Feynman propagator described by
\begin{equation}
\Delta (x-y)
= \int \frac{d^d k}{(2 \pi)^d} \,
\frac{e^{ik \, (x-y)}}{k^2+M^2-i \epsilon}
\equiv\int \frac{d^d k}{(2 \pi)^d} \,
e^{ik \, (x-y)}\,\tilde{\Delta}(k)\,.
\end{equation}
The property of the Feynman propagator
\begin{equation}
	(-\partial_{x}^{2}+M^{2})\,\Delta(x-y)=\delta^{d}(x-y)
\end{equation}
guarantees to satisfy~\eqref{HK}. 

To illustrate the formula for correlation functions, we define
\begin{equation}
{\bf U} = \int d^d x \, {\bm c} (x) \, {\bm d} (x) \,, \label{U-scalar}
\end{equation}
where ${\bm c} (x)$ and ${\bm d} (x)$ are coderivations with
\begin{equation}
\pi_1 \, {\bm c} (x) \, {\bf 1} = c(x) \,, \quad
\pi_1 \, {\bm c} (x) \, \pi_n = 0 \,, \quad
\pi_1 \, {\bm d} (x) \, {\bf 1} = d(x) \,, \quad
\pi_1 \, {\bm d} (x) \, \pi_n = 0
\end{equation}
for $n > 0 \,$.
Then, the formula for correlation functions in terms of quantum $A_{\infty}$ algebras are given by
\begin{equation}
\langle \, \Phi^{\otimes n} \, \rangle = \pi_n \, {\bm f} \, {\bf 1} \,,
\label{scalar-correlation-functions}
\end{equation}
where
\begin{equation}
\Phi^{\otimes n} = \underbrace{\, \Phi \otimes \Phi \otimes \ldots \otimes \Phi \,}_n
\end{equation}
and
\begin{equation}
\begin{split}
{\bm f} &= \frac{1}{{\bf I} +{\bm h} \, {\bm m} +i \hbar \, {\bm h} \, {\bf U}} \\
	&= {\bf I} +\sum_{n=1}^\infty \, (-1)^n \,
( \, {\bm h} \, {\bm m} +i \hbar \, {\bm h} \, {\bf U} \, )^n \,.
\end{split}
\end{equation}
Since
\begin{equation}
\begin{split}
\langle \, \Phi^{\otimes n} \, \rangle
& = \langle \, \underbrace{\, \Phi \otimes \Phi \otimes \ldots \otimes \Phi \,}_n \, \rangle \\
& = \int d^d x_1 d^d x_2 \ldots d^d x_n \,
\langle \, \varphi (x_1) \, \varphi (x_2) \, \ldots \, \varphi (x_n) \, \rangle \,
c (x_1) \otimes c(x_2) \otimes \ldots \otimes c(x_n) \,,
\end{split}
\end{equation}
the formula~\eqref{scalar-correlation-functions} can be rewritten as
\begin{equation}
\pi_n \, {\bm f} \, {\bf 1}
= \int d^d x_1 d^d x_2 \ldots d^d x_n \,
\langle \, \varphi (x_1) \, \varphi (x_2) \, \ldots \, \varphi (x_n) \, \rangle \,
c (x_1) \otimes c(x_2) \otimes \ldots \otimes c(x_n) \,.
\end{equation}
The definition of the symplectic form
\begin{equation}
\omega \, ( \, c(x) \,, d(x') \, ) = \delta^d (x-x') 
\end{equation}
enables us to extract the correlation function by 
\begin{equation}
\langle \, \varphi (x_1) \, \varphi (x_2) \, \ldots \, \varphi (x_n) \, \rangle
= \omega_n \, ( \, \pi_n \, {\bm f} \, {\bf 1} \,,
d (x_1) \otimes d (x_2) \otimes \ldots \otimes d (x_n) \, ) \,, \label{ex-scalar}
\end{equation}
where
\begin{equation}
{\bm f} = \frac{1}{{\bf I} +{\bm h} \, {\bm m} +i \hbar \, {\bm h} \, {\bf U}} \,,
\end{equation}
and
\begin{equation}
\omega_n \, ( \, c (x_1)  \otimes c (x_2) \otimes \ldots \otimes c (x_n) \,,
d (x'_1) \otimes d (x'_2) \otimes \ldots \otimes d (x'_n) \, )
= \prod_{i=1}^n \, \omega \, ( \, c (x_i) \,, d (x'_i) \, ) \,.
\label{omega_n-c-d}
\end{equation}

Let us calculate the two-point functions of the free theory. The main quantity we have to calculate is $ \pi_{2} \, {\bm f} \, {\bf 1}$. Since
\begin{equation}
\begin{split}
\, \pi_{2} \, {\bm f} \, {\bf 1} \,
&= -i \, \hbar \, \pi_{2} \, {\bm h} \, {\bf U} \, {\bf 1}\\
&= -i \, \hbar \, \int d^d x \,\int d^d y \,[\,c(x) \,\otimes\Delta(x-y)\,c(y) \,]\,,
\end{split}
\end{equation}
we obtain
\begin{equation}
      \omega_2 \, ( \, \pi_2 \, {\bm f} \, {\bf 1} \,,d (x_1) \otimes d (x_2) \, ) \, = \,\frac{\hbar}{i}\,\Delta\,(x_1\,-\,x_2\,)\,= \langle \, \varphi (x_1) \, \varphi (x_2) \,  \rangle
\end{equation}
to find that this calculation correctly reproduces the two-point functions of the free theory.

The correlation functions from the above formula satisfy the Schwinger-Dyson equations. The proof can be done by the identity
\begin{equation}
	{\bm f}^{-1}\,{\bm f}\,{\bf 1}=0\,.
\end{equation}
For more examples of concrete calculations and the proof of the Schwinger-Dyson equations in the free theory, see  \cite{Konosu/Okawa_2022}.
Thus, the ``correlation functions'' on the right-hand side of~\eqref{ex-scalar} describe the ordinary correlation functions in quantum field theory.
 
\subsection{Dirac fields}\label{Dirac-section}
In this subsection, we consider free Dirac fields in $d$ dimensions. Again, we consider the vector space
\begin{equation}
	\mathcal{H}=\mathcal{H}_{1}\oplus\mathcal{H}_{2}\,.
\end{equation}
The element $\Phi$ of $\mathcal{H}_{1}$ can be expanded as
\begin{equation}
\Phi = \int d^d x \, ( \, \overline{\theta}_\alpha (x) \, \Psi_\alpha (x)
+\overline{\Psi}_\alpha (x) \, \theta_\alpha (x) \, ) \,, \label{Dirac-expansion}
\end{equation}
where $\Psi_{\alpha} (x)$ is a Dirac field and $\theta_{\alpha}(x)$ is the basis vectors of $\mathcal{H}_{1}$. The ``bars'' on $\Psi_{\alpha} (x)$ and $\theta_{\alpha}(x)$ represent their Dirac adjoint.
 We also define $\Psi (x)$, $\overline{\Psi} (x)$, $\theta (x)$, and $\overline{\theta} (x)$ as degree odd.
 
To describe the action in terms of $A_{\infty}$ algebras, let us define the symplectic form $\omega$.
We define the non-zero symplectic form $\omega$ by
\begin{align}
\omega \, ( \, \theta_{\alpha_1} (x_1) \,, \overline{\lambda}_{\alpha_2} (x_2) \, )
& = \delta_{\alpha_1 \alpha_2} \, \delta^d ( x_1-x_2 ) \,, \\ 
\omega \, ( \, \overline{\theta}_{\alpha_1} (x_1) \,, \lambda_{\alpha_2} (x_2) \, ) 
& = \delta_{\alpha_1 \alpha_2} \, \delta^d ( x_1-x_2 ) \,, \\ 
\omega \, ( \,  \overline{\lambda}_{\alpha_2} (x_2)\,, \theta_{\alpha_1}(x_1)  \, ) 
& = {}-\delta_{\alpha_1 \alpha_2} \, \delta^d ( x_1-x_2 ) \,, \\
\omega \, ( \, \lambda_{\alpha_2} (x_2)  \,, \overline{\theta}_{\alpha_1} (x_1)\, )
& = {}-\delta_{\alpha_1 \alpha_2} \, \delta^d ( x_1-x_2 ) \,,
\end{align}
where  $\lambda_\alpha (x)$ and its Dirac adjoint $\overline{\lambda}_\alpha (x)$ are the basis vectors for the vector space $\mathcal{H}_2$, which is degree even. We then define the operator $Q$, which is degree-odd by
\begin{equation}
\begin{split}
& Q \, \theta_\alpha (x)
= {}( {}-i \, \partial\!\!\!/ +m \, )_{\alpha \beta} \, \lambda_\beta (x) \,, \quad
Q \, \lambda_\alpha (x) = 0 \,, \\
& Q \, \overline{\theta}_\alpha (x)
= {}-\overline{\lambda}_\beta (x) \, ( \, i \overleftarrow{\partial\!\!\!/} +m \, )_{\beta \alpha} \,, \quad
Q \, \overline{\lambda}_\alpha (x) = 0 \,,
\end{split}
\end{equation}
where $m$ is the mass of the Dirac field to obtain
\begin{equation}
	\begin{split}
		S &={}-\frac{1}{2} \, \omega \, ( \, \Phi \,,\, Q \, \Phi \, )\\
		   &= \,\int d^d x \, \bigl[ \, i \, \overline{\Psi} (x) \, \partial\!\!\!/ \, \Psi (x)-m \, \overline{\Psi} (x) \, \Psi (x) \, \bigr]
 	\end{split}
\end{equation}
with $\Phi$ in~\eqref{Dirac-expansion}. In fact, we need careful treatment when we deal with degree-odd fields. Let us consider $\omega \, ( \, \Phi_1 \,,\, \Phi_2 \, )$, where
 $\Phi_1$ and $\Phi_2$ are given by
\begin{equation}
\Phi_1 = f_1 \, e_1 \,, \qquad \Phi_2 = f_2 \, e_2 \,.
\label{Phi-form}
\end{equation}
In the process of taking out $f_1$ and $f_2$ from
$\omega \, ( \, f_1 \, e_1 \,,\, f_2 \, e_2 \, )$, the sign factors may arise. In  \cite{Konosu/Okawa_2022}, we discuss the subtlety of defining this sign factor and  we use the convention
\begin{equation}
\omega \, ( \, f_1 \, e_1 \,,\, f_2 \, e_2 \, )
= (-1)^{\mathrm{deg} (f_1)+\mathrm{deg} (f_2)+\mathrm{deg} (e_1) \mathrm{deg} (f_2)} \,
f_1 \, f_2 \, \omega \, ( \, e_1 \,,\, e_2 \, ) \,.
\label{omega-convention}
\end{equation}

Since the operator $Q$ satisfies the cyclicity and the nilpotency, we obtain  free Dirac theory in terms of $A_{\infty}$ algebras. Let us express the formula for correlation functions in quantum $A_{\infty}$ algebras. Again, we take the projection $P$ as
\begin{equation}
	P = 0\,.
\end{equation}
Then,  the definitions~\eqref{HK}  become
\begin{equation}
Q \, h +h \, Q = \mathbb{I} \,, \qquad
h^2 = 0 \,.
\end{equation}
As in the case  of scalar field theory, we use the Feynman propagator
\begin{equation}
S(x-y)_{\alpha \beta}
= \int \frac{d^d p}{(2 \pi)^d} \, e^{ip \, (x-y)} \,
\frac{( {}-p\!\!\!/ +m \, )_{\alpha \beta}}{p^2 +m^2 -i \epsilon}
\equiv \int \frac{d^d p}{(2 \pi)^d} \, e^{ip \, (x-y)} \,
\tilde{S}_{\alpha\beta}(p)\,,
\end{equation}
and we define
\begin{equation}
\begin{split}
& h \, \theta_\alpha (x) = 0 \,, \qquad
h \, \lambda_\alpha (x) = {}\int d^d y \, S(x-y)_{\alpha \beta} \, \theta_\beta (y) \,, \\
& h \, \overline{\theta}_\alpha (x) = 0 \,, \qquad
h \, \overline{\lambda}_\alpha (x)
= -\int d^d y \, \overline{\theta}_\beta (y) \, S(y-x)_{\beta \alpha} \,.
\end{split}
\label{Dirac-h}
\end{equation}

To illustrate the formula for correlation functions, we define the operator ${\bf U}$ by
\begin{equation}
{\bf U} = -\int d^d x \, ( \, \overline{\bm \theta}_\alpha (x) \, {\bm \lambda}_\alpha (x)
+{\bm \theta}_\alpha (x) \, \overline{\bm \lambda}_\alpha (x) \, ) \,, \label{U-Dirac}
\end{equation}
where ${\bm \theta}_\alpha (x),\,\overline{\bm \theta}_\alpha (x),\,{\bm \lambda}_\alpha (x)$ and $\overline{\bm \lambda}_\alpha (x)$ are
coderivations with
\begin{equation}
\pi_1 \, {\bm \theta}_\alpha (x) \, {\bf 1} = \theta_\alpha (x) \,, \quad
\pi_1 \, {\bm \theta}_\alpha (x) \, \pi_n = 0 \,, \quad
\pi_1 \, \overline{\bm \theta}_\alpha (x) \, {\bf 1} = \overline{\theta}_\alpha (x) \,, \quad
\pi_1 \, \overline{\bm \theta}_\alpha (x) \, \pi_n = 0
\end{equation}
\begin{equation}
\pi_1 \, {\bm \lambda}_\alpha (x) \, {\bf 1} = \lambda_\alpha (x) \,, \quad
\pi_1 \, {\bm \lambda}_\alpha (x) \, \pi_n = 0 \,, \quad
\pi_1 \, \overline{\bm \lambda}_\alpha (x) \, {\bf 1} = \overline{\lambda}_\alpha (x) \,, \quad
\pi_1 \, \overline{\bm \lambda}_\alpha (x) \, \pi_n = 0
\end{equation}
for $n > 0 \,$.

Then, the formula for correlation functions is given by
\begin{equation}
\langle \, \Phi^{\otimes n} \, \rangle = \pi_n \, {\bm f} \, {\bf 1} \,,
\label{Dirac-correlation-functions}
\end{equation}
where
\begin{equation}
\Phi^{\otimes n} = \underbrace{\, \Phi \otimes \Phi \otimes \ldots \otimes \Phi \,}_n
\end{equation}
and
\begin{equation}
{\bm f} = \frac{1}{{\bf I} +{\bm h} \, {\bm m} +i \hbar \, {\bm h} \, {\bf U}} \,.
\end{equation}
In the same way as scalar field theory, we can extract the correlation functions from~\eqref{Dirac-correlation-functions}. In particular, we obtain
\begin{equation}
\begin{split}
& \langle \, \Psi_{\alpha_1} (x_1) \, \ldots \Psi_{\alpha_n} (x_n) \,
\overline{\Psi}_{\beta_1} (y_1) \, \ldots \overline{\Psi}_{\beta_m} (y_n) \, \rangle \\
& = \omega_{2n} \, ( \, \pi_{2n} \, {\bm f} \, {\bf 1} \,,
\lambda_{\alpha_1} (x_1) \otimes \ldots \otimes \lambda_{\alpha_n} (x_n)
\otimes \overline{\lambda}_{\beta_1} (y_1) \otimes \ldots \otimes \overline{\lambda}_{\beta_n} (y_n) \, ) \label{ex-Dirac}
\end{split}
\end{equation}
where
\begin{equation}
{\bm f} = \frac{1}{{\bf I} +{\bm h} \, {\bm m} +i \hbar \, {\bm h} \, {\bf U}} \,.
\end{equation}
Here, we introduce a slightly modified notation for $\omega_{n}$ from~\eqref{omega_n-c-d} as discussed in~\cite{Konosu/Okawa_2022}:
\begin{equation}
\begin{split}
&\omega_n \, ( \, \Phi_1 \otimes \Phi_2 \otimes \ldots \otimes \Phi_n \,,
\widetilde{\Phi}_1 \otimes \widetilde{\Phi}_2 \otimes \ldots \otimes \widetilde{\Phi}_n \, ) \\
& = (-1)^\sigma \, \omega \, ( \, \Phi_1 \,, \widetilde{\Phi}_1 \, ) \,
\omega \, ( \, \Phi_2 \,, \widetilde{\Phi}_2 \, ) \ldots \,
\omega \, ( \, \Phi_n \,, \widetilde{\Phi}_n \, ) \,,
\end{split}
\label{omega_n}
\end{equation}
where
\begin{equation}
\sigma
= \sum_{i=1}^{n-1} \sum_{j=i+1}^n \mathrm{deg} (\widetilde{\Phi}_i) \, \mathrm{deg} (\Phi_j)
+\sum_{k=1}^{n-1} \, (n-k) \,
( \, \mathrm{deg} (\Phi_k) +\mathrm{deg} (\widetilde{\Phi}_k) -1 \, ) \mod 2 \,.
\end{equation}
The sign factor originated from $\sigma$ guarantees the well-definedness of $\omega_{n}$. In most of the cases, the degree of $\Phi$ and $\widetilde{\Phi}$ are given by
\begin{equation}
	 \mathrm{deg} (\Phi_j)=1\,,\qquad\mathrm{deg} (\widetilde{\Phi}_i)=0 \label{degree-pattern1}
\end{equation}
or
\begin{equation}
	 \mathrm{deg} (\Phi_j)=\mathrm{deg} (\widetilde{\Phi}_i)=0\,, \label{degree-pattern2}
\end{equation}
and $\sigma$ can be calculated with ease. When the degree is given by~\eqref{degree-pattern1}, $\sigma$ becomes
\begin{equation}
	\sigma = 0\,.
\end{equation}
In the case that the degree is given by ~\eqref{degree-pattern2}, we find
\begin{equation}
	\begin{split}
		\sigma&={}-\sum_{k=1}^{n-1} \, (n-k)\\
			  &={}-\frac{n\,(n-1)}{2}\mod 2 \,.
	\end{split}
\end{equation}

Let us calculate two-point function for the free theory. We can calculate as
\begin{equation}
\begin{split}
\, \pi_{2} \, {\bm f} \, {\bf 1} \,
&= -i \, \hbar \, \pi_{2} \, {\bm h} \, {\bf U} \, {\bf 1}\\
&=-i \, \hbar \, \int d^d x \,\int d^d y \,[\,\overline{\theta}_{\alpha}(x)\, \otimes S_{\alpha \beta}(x-y)\, \theta_{\beta}(y)\,
-\theta_{\alpha}(x)\, \otimes \overline{\theta}_{\beta}(y)\,S_{\beta\alpha }(y-x)\, ]\,. 
\end{split}
\end{equation}
When we extract the correlation function, we need to calculate
\begin{equation}
	\omega_2 \, ( \, \pi_2 \, {\bm f} \, {\bf 1} \,,\lambda_{\alpha_1} (x_1) \otimes \overline{\lambda}_{\beta_1} (y_1) \, ) \,.
\end{equation}
In this case, the degree inside $\omega_{2}$ corresponds to the case~\eqref{degree-pattern1} and we obtain
\begin{equation}
      \omega_2 \, ( \, \pi_2 \, {\bm f} \, {\bf 1} \,,\lambda_{\alpha_1} (x_1) \otimes \overline{\lambda}_{\beta_1} (y_1) \, ) \, = \,\frac{\hbar}{i}\,S_{\alpha_1\beta_1}\,(x_1\,-\,y_1\,)\,= \langle \, \Psi_{\alpha_1} (x_1) \, \overline{\Psi}_{\beta_1} (y_1) \,  \rangle\,,
\end{equation}
and this calculation reproduce the known result. It is also possible to calculate $\langle \,  \overline{\Psi}_{\beta_1} (y_1) \, \Psi_{\alpha_1} (x_1) \, \rangle$:
\begin{equation}
      \omega_2 \, ( \, \pi_2 \, {\bm f} \, {\bf 1} \,, \overline{\lambda}_{\beta_1} (y_1)\otimes \lambda_{\alpha_1} (x_1) \, ) \, = {}-\frac{\hbar}{i}\,S_{\alpha_1\beta_1}\,(x_1\,-\,y_1\,)\,= \langle \,  \overline{\Psi}_{\beta_1} (y_1) \, \Psi_{\alpha_1} (x_1) \, \rangle\,.
\end{equation}

Again, the correlation functions from the above formula satisfy the Schwinger-Dyson equations. Thus, the ``correlation functions'' on the right-hand side of~\eqref{ex-Dirac} describe the ordinary correlation functions in quantum field theory. For examples of concrete calculations and the proof of the Schwinger-Dyson equations for free theory, see~\cite{Konosu/Okawa_2022}.

\section{Modified Yukawa theory}\label{mod-Yukawa}
In this section, we consider loop corrections of the modified Yukawa theory in $d=4$ dimensions and confirm that we can express the  formula for correlation functions presented in~\cite{Konosu/Okawa_2022} in the same form.
\setcounter{equation}{0}
\subsection{Action}
First, we express the Yukawa theory in terms of $A_{\infty}$ algebras.
For simplicity, we modify the Yukawa interaction and require parity symmetry. Since the field $\varphi(x)$ becomes a pseudoscalar field, we can omit the $\varphi^3$ term as scalar interactions. 
We consider an action
\begin{equation}
  S = \int \,d^4 x\, (\mathcal{L}_0 \,+ \,\mathcal{L}_1),\label{mod-yukawa}
\end{equation}
where
\begin{equation}
  \mathcal{L}_0 = -\frac{1}{2}\,\partial^{\mu}\varphi(x) \partial_{\mu}\varphi(x) \,-\frac{1}{2}\,M^2\varphi(x)^2\,+i\overline{\Psi}(x){}\partial\!\!\!/\Psi(x)-m\overline{\Psi}(x)\Psi(x)\,,
\end{equation}
\begin{equation}
  \mathcal{L}_1 = i\,Z_g \, g\, \varphi(x)\, \overline{\Psi}(x)\,\gamma_5\,\Psi(x) -\,\frac{1}{24}\, Z_{\lambda}\, \lambda \,\varphi(x)^4 \,+ \,\mathcal{L}_{\mathrm{ct}}\,,
\end{equation}
and
\begin{equation}
  \begin{split}
  \mathcal{L}_{\mathrm{ct}} = &-\frac{1}{2}\,(Z_{\varphi}-1)\partial^{\mu}\varphi(x) \partial_{\mu}\varphi(x) \,-\frac{1}{2}\,(Z_{M}\,-\,1)\,M^2\varphi^2\,\\
  &+i\,(Z_{\Psi}\,-1)\overline{\Psi}(x){}\partial\!\!\!/\Psi(x)-\,(Z_m\,-\,1)m\overline{\Psi}(x)\Psi(x)\,.
\end{split}
\end{equation}
We call this theory the ``\textit{modified Yukawa theory}'' to distinguish from the original Yukawa theory.\footnote{Discussions of the modified Yukawa theory in the path integral formalism are presented, for example, in section 51 of~\cite{Srednicki:2007qs}.}
Again, we consider the vector space
\begin{equation}
	\mathcal{H}=\mathcal{H}_{1}\oplus\mathcal{H}_{2}\,,
\end{equation}
and we rewrite this action to
\begin{equation}
S = {}-\frac{1}{2} \, \omega \, ( \, \Phi, Q \, \Phi \, )
-\sum_{n=0}^{3} \, \frac{1}{n+1} \,
\omega \, ( \, \Phi \,, m_n \, ( \, \Phi \otimes \ldots \otimes \Phi \, ) \, ) \,, \label{hom-mod-Yukawa}
\end{equation}
where $\Phi$ in $\mathcal{H}_1$ is expanded as
\begin{equation}
\Phi = \int d^4 x \left[\, \varphi (x) \, c(x) + \, \overline{\theta}_\alpha (x) \, \Psi_\alpha (x)
+\overline{\Psi}_\alpha (x) \, \theta_\alpha (x) \, \right] \,.
\end{equation}
The definitions of $\omega,\,Q,\,h$ on basis vectors $c(x),\, d(x),\, \theta(x),\, \overline{\theta}(x),\, \lambda(x)$, and $\overline{\lambda}(x)$ are same as section \ref{review-section}. 
We define non-zero $m_1$ as
\begin{equation}
  \begin{split}
    m_1\,c(x) &= [-(Z_{\varphi}\,-\,1)\,\partial^2\,+\,(Z_M\,-\,1)\,M^2]\,d(x)\,,\\
    m_1\,\theta_{\alpha}(x)\,&=[-\,i\,(Z_{\Psi}\,-\,1\,)\,{}\partial\!\!\!/\,+\,(Z_m\,-\,1\,)\,m\,]_{\alpha \beta}\,\lambda_{\beta}(x)\,,\\
    m_1\,\overline{\theta}_{\alpha}(x)\, &= \,-\,\overline{\lambda}_{\beta}(x)\,[\,i\,(Z_{\Psi}\,-\,1\,)\,\overleftarrow{\partial\!\!\!/}\,+\,(Z_m\,-\,1\,)\,m\,]_{\beta\alpha}\,.
  \end{split}
\end{equation}
We find that these definitions reproduce the counterterms in the action~\eqref{mod-yukawa}:
\begin{equation}
  -\frac{1}{2}\,\omega(\,\Phi,\,m_1(\Phi))= \int d^4 x \,\mathcal{L}_{ct}\,.
\end{equation}
We define non-zero $m_2$ as
\begin{equation}
  \begin{split}
    m_2(\theta_{\alpha}(x_1)\,\otimes\,\overline{\theta}_{\beta}(x_2))\,&=\,-\frac{1}{2}\,i\,Z_g\,g\,(\gamma_5)_{\alpha\beta}\,\delta^4(x_1-x_2)\,d(x_1), \\
    m_2(\overline{\theta}_{\alpha}(x_1)\,\otimes\,\theta_{\beta}(x_2))\,&=\,\,\,\,\,\,\frac{1}{2}\,i\,Z_g\,g\,(\gamma_5)_{\beta\alpha}\,\delta^4(x_1-x_2)\,d(x_1), \\
    m_2(\,c\,(x_1)\,\otimes\,\theta_{\alpha}(x_2))\,\,\,&=\,-\frac{1}{2}\,i\,Z_g\,g\,(\gamma_5)_{\alpha\gamma}\,\lambda_{\gamma}(x_1)\,\delta^4(x_1-x_2)\,, \\
    m_2(\theta_{\alpha}(x_1)\otimes\,\,c\,(x_2)\,)\,\,\,&=\,-\frac{1}{2}\,i\,Z_g\,g\,(\gamma_5)_{\alpha\gamma}\,\lambda_{\gamma}(x_1)\delta^4(x_1-x_2)\,, \\
    m_2(\overline{\theta}_{\alpha}(x_1)\otimes\,\,c\,(x_2)\,)\,\,\,&=\,\,\,\,\,\,\frac{1}{2}\,i\,Z_g\,g\,\delta^4(x_1-x_2)\,\overline{\lambda}_{\gamma}(x_1)(\gamma_5)_{\gamma\alpha}\,, \\
    m_2(\,c\,(x_1)\,\otimes\,\overline{\theta}_{\alpha}(x_2))\,\,\,&=\,\,\,\,\,\,\frac{1}{2}\,i\,Z_g\,g\,\delta^4(x_1-x_2)\,\overline{\lambda}_{\gamma}(x_1)(\gamma_5)_{\gamma\alpha}\,. \\
  \end{split}
\end{equation}
We find that these definitions reproduce the modified Yukawa interaction in the action~\eqref{mod-yukawa}:
\begin{equation}
  -\frac{1}{3}\,\omega(\,\Phi,\,m_2(\Phi\otimes\Phi))= i\,Z_g \, g\, \int d^4 x \,\varphi(x)\, \overline{\Psi}(x)\,\gamma_5\,\Psi(x)\,.
\end{equation}
We define non-zero $m_3$ as
\begin{equation}
  m_3(\,c(x_1)\,\otimes\, c(x_2)\,\otimes\, c(x_3)\,) \,=\, \frac{1}{6}\,Z_{\lambda}\,\lambda\,\delta^4(\,x_1\,-\,x_2\,)\,\delta^4(\,x_2\,-\,x_3\,)\,d(x_1).
\end{equation}
We find that this definition reproduces the scalar $\varphi^{4}$ interaction in the action~\eqref{mod-yukawa}:
\begin{equation}
  -\frac{1}{4}\,\omega(\,\Phi,\,m_3(\Phi\,\otimes\,\Phi\,\otimes\,\Phi))= -\frac{1}{24}\, Z_{\lambda}\, \lambda \,\int d^4 x \,\,\varphi(x)^4 \,.
\end{equation}
Then, we have rewritten the actions~\eqref{mod-yukawa} to the action of the form~\eqref{hom-mod-Yukawa} in terms of quantum $A_{\infty}$ algebras.

\subsection{Correlation functions}
In this subsection, we confirm that the formula for correlation functions in terms of quantum $A_{\infty}$ algebras for the modified Yukawa theory takes the same form as~\eqref{scalar-correlation-functions} and~\eqref{Dirac-correlation-functions} through calculating several one-loop correlation functions explicitly. We claim that correlation functions for the modified Yukawa theory are obtained by
\begin{equation}
\langle \, \Phi^{\otimes n} \, \rangle = \pi_n \, {\bm f} \, {\bf 1} \,,
\end{equation}
where
\begin{equation}
\Phi = \int d^4 x \left[\, \varphi (x) \, c(x) + \, \overline{\theta}_\alpha (x) \, \Psi_\alpha (x)
+\overline{\Psi}_\alpha (x) \, \theta_\alpha (x) \, \right] \,,
\end{equation}
\begin{equation}
{\bm f} = \frac{1}{{\bf I} +{\bm h} \, {\bm m} +i \hbar \, {\bm h} \, {\bf U}} \,.
\end{equation}
For the particular correlation functions,  we can rewrite the above formula as
\begin{equation}
 	\begin{split}
 		 &\langle \, \Psi_{\alpha_1} (y_1) \, \ldots \Psi_{\alpha_m} (y_m) \,
  				\overline{\Psi}_{\beta_1} (z_1) \, \ldots \overline{\Psi}_{\beta_{m}} (z_{m}) \, \varphi(x_{1})\,\ldots\,\varphi(x_{n})\, \rangle\,\\
				&=\omega_{2m+n} \, ( \, \pi_{2m+n}\,{\bm{f}}\,{\bf 1}\,\,,\\
			&\quad\lambda_{\alpha_1} (y_1) \otimes \ldots \otimes \lambda_{\alpha_m} (y_m)
 			 \otimes \overline{\lambda}_{\beta_1} (z_1) \otimes \ldots \otimes \overline{\lambda}_{\beta_{l}} (z_{l}) \otimes d(x_{1})\otimes\ldots\otimes d(x_{n})\, )\,. \label{Yukawa-cor}
	\end{split}
\end{equation}
To calculate correlation functions, we define
\begin{equation}
{\bf U} = \int d^4 x  \, \left[{\bm c} (x) \, {\bm d} (x)-( \, \overline{\bm \theta}_\alpha (x) \, {\bm \lambda}_\alpha (x)
+{\bm \theta}_\alpha (x) \, \overline{\bm \lambda}_\alpha (x) \, )\right] \,.
\end{equation}
This definition is the simple sum of~\eqref{U-scalar} and~\eqref{U-Dirac}.

First, let us calculate the one-loop corrections of the one-point functions. The key ingredient to calculate the one-point functions is $\pi_{1}\,{\bm f}\,{\bf 1}$:
\begin{equation}
	\begin{split}
		\pi_{1}\,{\bm f}\,{\bf 1} &= \pi_{1}\,\frac{1}{{\bf I}+{\bm h}\,{\bm m}+i\hbar\,{\bm h}\,{\bf U}}\,{\bf 1}\\
					      &=i\hbar\,\pi_{1}{\bm h}\,{\bm m}_{2}\,{\bm h}\,{\bf U}\,{\bf 1}\,+\,\ldots\,.
	\end{split}
\end{equation}
The $\ldots$ are higher-order terms. Then, we obtain
\begin{equation}
	\pi_{1}\,{\bm h}\,{\bm m}_{2}\,{\bm h}\,{\bf U}\,{\bf 1} = h\,m_{2}\,\pi_{2}\,{\bm h}\,{\bf U}\,{\bf 1}\,.
\end{equation}
The term $\pi_{2}\,{\bm h}\,{\bf U}\,{\bf 1}$ can be easily calculated and the result is as follows:
\begin{equation}
	\pi_{2}\,{\bm h}{\bf U}\,{\bf 1} = \int \,d^{4}x\,\left[c(x)\otimes hd(x)+\overline{\theta}_{\alpha}\otimes h\lambda_{\alpha}(x)+\theta_{\alpha}(x)\otimes h\overline{\lambda}_{\alpha}(x)\right]\,. \label{hu}
\end{equation}
Then, we obtain the one-loop correlation functions of the Dirac field and its Dirac adjoint:
\begin {align}
        \langle \, \Psi_{\alpha_1} (x_1) \, \rangle^{(1)}&=  \omega_{1} \, ( \, \pi_{1} \, {\bm f} \, {\bf 1} \,, \lambda_{\alpha_1} (x_1) \, )=0 \,,\\
        \langle \, \overline{\Psi}_{\beta_1} (y_1) \, \rangle^{(1)}&=  \omega_{1} \, ( \, \pi_{1} \, {\bm f} \, {\bf 1} \,, \overline{\lambda}_{\beta_1} (y_1) \, )=0 \,, 
\end {align}
where $\langle\,\cdot\,\rangle^{(1)}$ is correlation function up to one-loop corrections.
The one-point function for the scalar field is nontrivial. From the calculation (\ref{hu}), we obtain
\begin{equation}
	\pi_{1}{\bm h}{\bm m}_{2}\,{\bm h}{\bf U}\,{\bf 1}\, = \frac{1}{2}\,i\,Z_{g}\,g\,\int \,d^{4}x\,\int\,d^{4}y\,\delta^{4}(x-y)\left\{\mathrm{Tr}[\gamma_{5}S(x-y)]+\mathrm{Tr}[\gamma_{5}S(y-x)]\right\}hd(x)\,.
\end{equation}
These integrals are divergent. As in ordinary formalisms of quantum field theory, we regularize them by using, for instance, the Pauli-Villars regularization and make them finite. Notice that the term $S(x-y)$ contains one gamma matrix. The term $\mathrm{Tr}[\gamma_{5}S(x-y)]$ contains $\mathrm{Tr}(\gamma_{5}\gamma_{\mu})$, which becomes zero. Then, we obtain the one-loop correlation functions of the scalar field:
\begin{equation}
	\langle \, \varphi(x)\, \rangle^{(1)} =  \omega_{1} \, ( \, \pi_{1} \, {\bm f} \, {\bf 1} \,, d(x) \, )=0 \,.
\end{equation}

\begin{figure}[htbp]    
\begin{center}      
\includegraphics[width=8 cm]{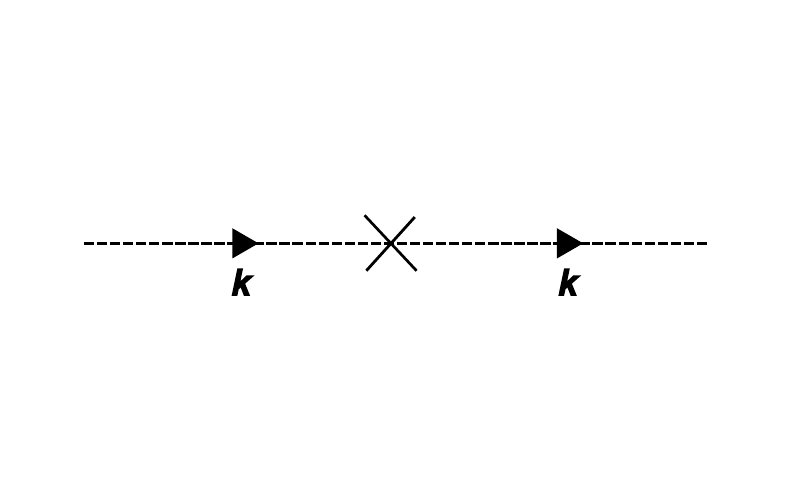}
\caption{the scalar counterterm correction}
\label{scalar-ct}
\end{center}
\end{figure}

Next, let us consider two-point correlation functions. The key ingredient to calculate the two-point functions is $\pi_{2}\,{\bm f}\,{\bf 1}$. This can be expanded as follows:
\begin{equation}
	\begin{split}
	\pi_{2}\,{\bm f}\,{\bf 1} &= \pi_{2}\,\frac{1}{{\bf I}+{\bm h}\,{\bm m}+i\hbar\,{\bm h}\,{\bf U}}\,{\bf 1}\\
				      &= \pi_{2}\,\frac{1}{{\bf I}+i\hbar\,{\bm h}\,{\bf U}}\,{\bf 1}-\pi_{2}\,\frac{1}{{\bf I}+i\hbar\,{\bm h}\,{\bf U}}\,{\bm h}\,{\bm m}\,\frac{1}{{\bf I}+i\hbar\,{\bm h}\,{\bf U}}\,{\bf 1}\\
				      &\quad+\pi_{2}\,\frac{1}{{\bf I}+i\hbar\,{\bm h}\,{\bf U}}\,{\bf h}\,{\bm m}\,\frac{1}{{\bf I}+i\hbar\,{\bm h}\,{\bf U}}\,{\bm h}\,{\bm m}\,\frac{1}{{\bf I}+i\hbar\,{\bm h}\,{\bf U}}\,{\bf 1}+\ldots
	\end{split}
\end{equation}
Since
\begin{equation}
	 \pi_{2}\,\frac{1}{{\bf I}+i\hbar\,{\bm h}\,{\bf U}}=\pi_{2}-i\hbar\,\pi_{2}\,{\bm h}\,{\bf U}\,\pi_{0}\,,
\end{equation}
we obtain
\begin{equation}
	\pi_{2}\,{\bm f}\,{\bf 1} = -i\hbar\,\pi_{2}\,{\bm h}\,{\bf U}\,{\bf 1}-\pi_{2}\,{\bm h}\,{\bm m}\,\frac{1}{{\bf I}+i\hbar\,{\bm h}\,{\bf U}}\,{\bf 1}+\pi_{2}\,{\bm h}\,{\bm m}\,\frac{1}{{\bf I}+i\hbar\,{\bm h}\,{\bf U}}\,{\bm h}\,{\bm m}\,\frac{1}{{\bf I}+i\hbar\,{\bm h}\,{\bf U}}\,{\bf 1}+\ldots\,.\label{pi2f}
\end{equation}
The first term is the contribution of the free part. Let us consider the second term and the third term in (\ref{pi2f}).
The second term can be calculated as follows:
\begin{equation}
	\begin{split}
		-\pi_{2}\,{\bm h}\,{\bm m}\,\frac{1}{{\bf I}+i\hbar\,{\bm h}\,{\bf U}}\,{\bf 1}&=-\pi_{2}\,({\bm h}\,{\bm m}_{1}+{\bm h}\,{\bm m}_{2}+{\bm h}\,{\bm m}_{3})\left({\bf I}+\sum_{n=1}^{\infty}(-1)^{n}(i\hbar\,{\bm h}\,{\bf U})^{n}\right){\bf 1}\\
												    &=i\hbar\,\pi_{2}\,{\bm h}{\bm m}_{1}\,{\bm h}\,{\bf U}\,{\bf 1}-(i\hbar)^{2}\,\pi_{2}\,{\bm h}\,{\bm m}_{3}\,{\bm h}\,{\bf U}\,{\bm h}\,{\bf U}\,{\bf 1}\,.
	\end{split}
\end{equation}
The third term in (\ref{pi2f}) can be calculated as follows:
\begin{equation}
	\begin{split}
		\pi_{2}{\bm h}{\bm m}\frac{1}{{\bf I}+i\hbar{\bm h}{\bf U}}{\bm h}{\bm m}\frac{1}{{\bf I}+i\hbar{\bm h}{\bf U}}{\bf 1}&=(i\hbar)^{2}\pi_{2}{\bm h}{\bm m}_{2}{\bm h}{\bm m}_{2}{\bm h}{\bf U}{\bm h}{\bf U}{\bf 1}+(i\hbar)^{2}\pi_{2}{\bm h}{\bm m}_{2}{\bm h}{\bf U}{\bm h}{\bm m}_{2}{\bm h}{\bf U}{\bf 1}\\
																				&\quad-i\hbar\pi_{2}{\bm h}{\bm m}_{1}{\bm h}{\bm m}_{1}{\bm h}{\bf U}{\bf 1}+(i\hbar)^{2}\pi_{2}{\bm h}{\bm m}_{1}{\bm h}{\bm m}_{3}{\bm h}{\bf U}{\bm h}{\bf U}{\bf 1}\\
																				&\quad+(i\hbar)^{2}\pi_{2}{\bm h}{\bm m}_{3}{\bm h}{\bm m}_{1}{\bm h}{\bf U}{\bm h}{\bf U}{\bf 1}+(i\hbar)^{2}\pi_{2}{\bm h}{\bm m}_{3}{\bm h}{\bf U}{\bm h}{\bm m}_{1}{\bm h}{\bf U}{\bf 1}\\
																				&\quad-(i\hbar)^{3}\pi_{2}{\bm h}{\bm m}_{3}{\bm h}{\bm m}_{3}{\bm h}{\bf U}{\bm h}{\bf U}{\bm h}{\bf U}{\bf 1}\\
																				&\quad-(i\hbar)^{3}\pi_{2}{\bm h}{\bm m}_{3}{\bm h}{\bf U}{\bm h}{\bm m}_{3}{\bm h}{\bf U}{\bm h}{\bf U}{\bf 1}\,. \label{third-pi2f}
	\end{split}
\end{equation}
Only the terms in the first line of the right-hand side in (\ref{third-pi2f}) contribute to the correlation functions. Note that
\begin{align}
	Z_{\varphi} &= 1+\mathcal{O}(g^{2})\,\\
	Z_{M}	&= 1+\mathcal{O}(\lambda,g^{2})\,\\
	Z_{\Psi}	&= 1+\mathcal{O}(g^{2})\,\\
	Z_{m} &= 1+\mathcal{O}(g^{2})\,.
\end{align}
Therefore, we do not need the terms in the second and the third lines of the right-hand side in (\ref{third-pi2f}) to compute the correction of the one-loop diagram.\footnote{These results follow from the renormalization, however, to simplify the situation, we assume the above results. Of course, we obtain the same result if you calculate all contributing terms without assuming the above.}  The terms in the fourth and fifth lines of the right-hand side in (\ref{third-pi2f}) are higher-order contributions. So, we need to calculate the following terms except for the first term which describes the contribution of the free part, and terms omitted by $\ldots$ which describes higher order terms:
\begin{equation}
	\begin{split}
	\pi_{2}\,{\bm f}\,{\bf 1}  &= -i\hbar\,\pi_{2}\,\frac{1}{{\bf I}+i\hbar\,{\bm h}\,{\bf U}}\,{\bf 1}+i\hbar\,\pi_{2}\,{\bm h}\,{\bm m}_{1}{\bm h}\,{\bf U}\,{\bf 1}-(i\hbar)^{2}\,\pi_{2}\,{\bm h}\,{\bm m}_{3}\,{\bm h}\,{\bf U}\,{\bm h}\,{\bf U}\,{\bf 1}\\
				       &\quad +(i\hbar)^{2}\,\pi_{2}\,{\bm h}\,{\bm m}_{2}\,{\bm h}\,{\bm m}_{2}\,{\bm h}\,{\bf U}\,{\bm h}\,{\bf U}{\bf 1}+(i\hbar)^{2}\,\pi_{2}\,{\bf h}\,{\bm m}_{2}\,{\bm h}\,{\bf U}\,{\bm h}\,{\bm m}_{2}\,{\bm h}\,{\bf U}\,{\bf 1}+\ldots \label{pi2f1-cal}
	\end{split}
\end{equation}Let us consider the second term in (\ref{pi2f1-cal}). This can be calculated as follows:
\begin{equation}
	i\hbar\,\pi_{2}\,{\bm h}\,{\bm m}_{1}\,{\bm h}\,{\bf U}\,{\bf 1} = i\hbar\int d^{4}x\,\left[c(x)\otimes hm_{1}hd(x)+\overline{\theta}_{\alpha}(x)\otimes hm_{1}h\lambda_{\alpha}(x)+\theta_{\alpha}(x)\otimes hm_{1}h\overline{\lambda}_{\alpha}(x)\right]\,. 
\end{equation}
Then, we obtain
\begin{equation}
	\omega_{2}(i\hbar\,\pi_{2}\,{\bm h}\,{\bm m}_{1}\,{\bm h}\,{\bf U}\,{\bf 1},d(x_{1})\otimes d(x_{2}))=i\hbar\int\frac{d^{4}k}{(2\pi)^{4}}\tilde{\Delta}(k)[(Z_{\varphi}-1)k^{2}+(Z_{M}-1)M^{2}]e^{ik(x_{1}-x_{2})}\tilde{\Delta}(k)\,.      
\end{equation}
This gives the contribution of the Fourier transform of the diagram of Figure.\ref{scalar-ct}.

\begin{figure}[t]    
\begin{center}      
\includegraphics[width=8 cm]{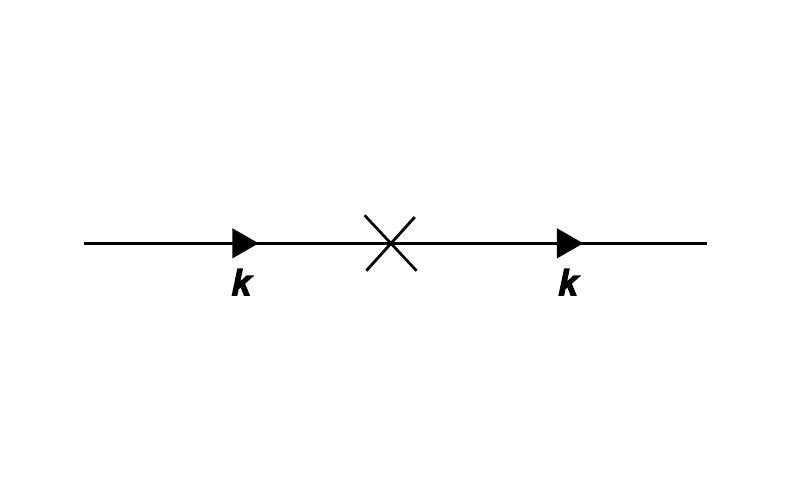}
\caption{the fermion counterterm correction}
\label{Dirac-ct}
\end{center}
\end{figure}

\noindent In the same way, we obtain
\begin{equation}
	\begin{split}
	&\omega_{2}(i\hbar\,\pi_{2}\,{\bm h}\,{\bm m}_{1}\,{\bm h}\,{\bf U}\,{\bf 1},\lambda_{\alpha_{1}}(x_{1})\otimes \overline{\lambda}_{\beta_{1}}(y_{1}))\\
	&\qquad=i\hbar\int\frac{d^{4}k}{(2\pi)^{4}}\tilde{S}(k)[(Z_{\Psi}-1){}k\!\!\!/\,+(Z_{m}-1)m]e^{ik(x_{1}-y_{1})}\tilde{S}(k)\,,  
	\end{split}    
\end{equation}
where we omit the spinor indices.
This gives the contribution of the Fourier transform of the diagram of Figure.\ref{Dirac-ct}.

\begin{figure}[htbp]   
\begin{center}      
\includegraphics[width=8 cm]{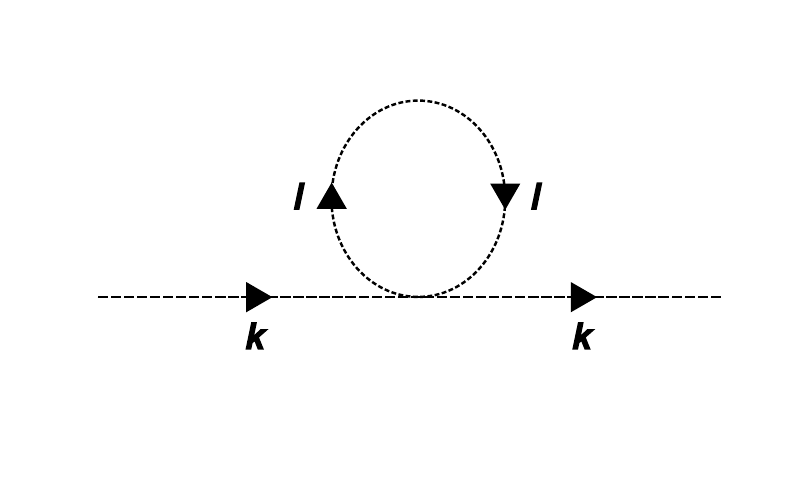}
\caption{the scalar 1-loop diagram}
\label{scalar-loop}
\end{center}
\end{figure}

\noindent Let us consider the third term in (\ref{pi2f1-cal}). This can be calculated as follows:
\begin{equation}
	\begin{split}
	&-(i\hbar)^{2}\,\pi_{2}\,{\bm h}\,{\bm m}_{3}\,{\bm h}\,{\bf U}\,{\bm h}\,{\bf U}\,{\bf 1}  \\
	&\qquad= \frac{1}{2}\hbar^{2}Z_{\lambda}\lambda\int d^{4}xd^{4}y\int\frac{d^{4}k}{(2\pi)^{4}}\frac{d^{4}k'}{(2\pi)^{4}}\tilde{\Delta}(k)\tilde{\Delta}(k')\tilde{\Delta}(k)e^{ik(x-y)}c(x)\otimes c(y)\,.
	\end{split}
\end{equation}
Then, we obtain
\begin{equation}
	\begin{split}
	&\omega_{2}(-(i\hbar)^{2}\,\pi_{2}\,{\bm h}\,{\bm m}_{3}\,{\bm h}\,{\bf U}\,{\bm h}\,{\bf U}\,{\bf 1},d(x_{1})\otimes d(x_{2}))\\
	&\qquad= \frac{1}{2}\hbar^{2}\lambda\int\frac{d^{4}k}{(2\pi)^{4}}e^{ik(x_{1}-k_{2})}\tilde{\Delta}(k)\Delta(0)\tilde{\Delta}(k)+\mathcal{O}(\lambda^{2})\,. 
	\end{split}   
\end{equation}
This gives the contribution of the Fourier transform of the diagram of Figure.\ref{scalar-loop}.

\begin{figure}[htbp]    
\begin{center}     
\includegraphics[width=10 cm]{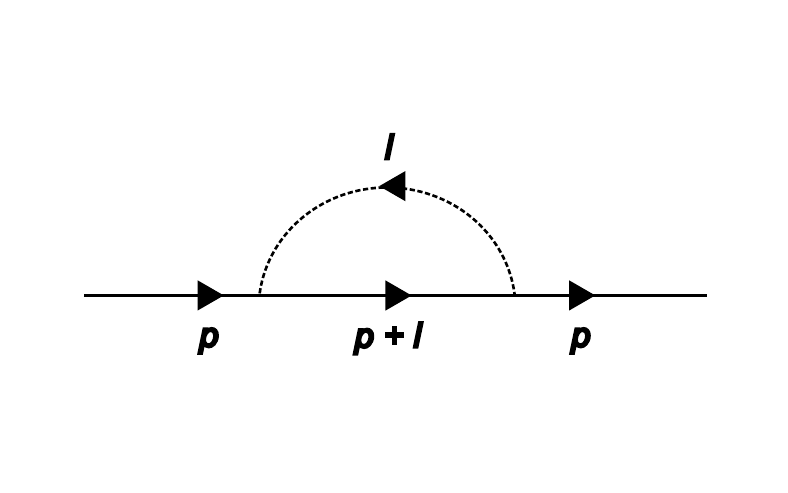}
\caption{the fermion diagram with scalar loop}
\label{fermion-scalar-loop}
\end{center}
\end{figure}

\noindent Let us consider the fourth term in (\ref{pi2f1-cal}). This can be calculated as follows:
\begin{equation}
	(i\hbar)^{2}\,\pi_{2}\,{\bm h}\,{\bm m}_{2}\,{\bm h}\,{\bm m}_{2}\,{\bm h}\,{\bf U}\,{\bm h}\,{\bf U}\,{\bf 1} = -\hbar^{2}\int d^{4}x d^{4}y \,\left[f(x,y)+g(x,y)+h(x,y)+i(x,y)\right]\,,
\end{equation}
where
\begin{equation}
	\begin{split}
		f(x,y)&=\overline{\theta}_{\beta}(y)\otimes hm_{2}(c(x)\otimes hm_{2}(hd(x)\otimes h\lambda_{\beta}(y)))\\
			&\quad+\overline{\theta}_{\alpha}(x)\otimes hm_{2}(c(y)\otimes hm_{2}(h\lambda_{\alpha}(x)\otimes hd(y) ))\\
			&\quad-\overline{\theta}_{\alpha}(x)\otimes hm_{2}(\theta_{\beta}(y)\otimes hm_{2}(h\lambda_{\alpha}(x)\otimes h\overline{\lambda}_{\beta}(y) ))\\
			&\quad+\overline{\theta}_{\beta}(y)\otimes hm_{2}(\theta_{\alpha}(x)\otimes hm_{2}(h\overline{\lambda}_{\alpha}(x)\otimes h\lambda_{\beta}(y)))\,,
	\end{split}
\end{equation}
\begin{equation}
	\begin{split}
		g(x,y)&=c(x)\otimes hm_{2}(\overline{\theta}_{\beta}(y)\otimes hm_{2}(hd(x)\otimes h\lambda_{\beta}(y)))\\
			&\quad+c(x)\otimes hm_{2}(\theta_{\beta}(y)\otimes hm_{2}(hd(x)\otimes h\overline{\lambda}_{\beta}(y)))\\
			&\quad+c(y)\otimes hm_{2}(\overline{\theta}_{\alpha}(x)\otimes hm_{2}(h\lambda_{\alpha}(x)\otimes hd(y)))\\
			&\quad+c(y)\otimes hm_{2}(\theta_{\alpha}(x)\otimes hm_{2}(h\overline{\lambda}_{\alpha}(x)\otimes hd(y)))\,,
	\end{split}
\end{equation}
and
\begin{equation}
	\begin{split}
		h(x,y)&=\overline{\theta}_{\alpha}(x)\otimes hm_{2}((h\lambda_{\alpha}(x)\otimes hm_{2}(\overline{\theta}_{\beta}(y)\otimes h\lambda_{\beta}(y)))\\
			&\quad+\overline{\theta}_{\alpha}(x)\otimes hm_{2}(h\lambda_{\alpha}(y)\otimes hm_{2}(\theta_{\beta}(y)\otimes h\overline{\lambda}_{\beta}(y)))\,.
	\end{split}
\end{equation}
As we will discuss later, the function $f(x,y)$ contributes to $ \langle  \Psi_{\alpha_1} (x_1)\overline{\Psi}_{\beta_1} (y_1)\rangle$ and $g(x,y)$ contributes to  $ \langle  \varphi (x_1)\varphi (x_2)\rangle$. The contribution $h(x,y)$ vanishes due to the regularization as in the case of scalar field theory.\footnote{Here, we do not write down $i(x,y)$ explicitly because this term is the contribution to \\$ \langle \overline{\Psi}_{\beta_1} (y_1) \Psi_{\alpha_1} (x_1) \rangle$, which we do not calculate here. } Then, we obtain 
\begin{equation}
	\begin{split}
		&\omega_{2}((i\hbar)^{2}\,\pi_{2}\,{\bm h}\,{\bm m}_{2}\,{\bm h}\,{\bm m}_{2}\,{\bm h}\,{\bf U}\,{\bm h}\,{\bf U}\,{\bf 1},\lambda_{\alpha_{1}}(x_{1})\otimes \overline{\lambda}_{\beta_{1}}(y_{1}))\\
		&=\hbar^{2}g^{2}\int\frac{d^{4}p}{(2\pi)^{4}}\frac{d^{4}l}{(2\pi)^{4}}e^{ip(x_{1}-y_{1})}\,\tilde{S}(p)\,\gamma_{5}\,\tilde{\Delta}(l)\,\tilde{S}(p+l)\,\gamma_{5}\,\tilde{S}(p)+\mathcal{O}(g^{3})\,.
	\end{split}
\end{equation}
This gives the contribution of the Fourier transform of the diagram of Figure.\ref{fermion-scalar-loop}.

\begin{figure}[t]    
\begin{center}      
\includegraphics[width=10 cm]{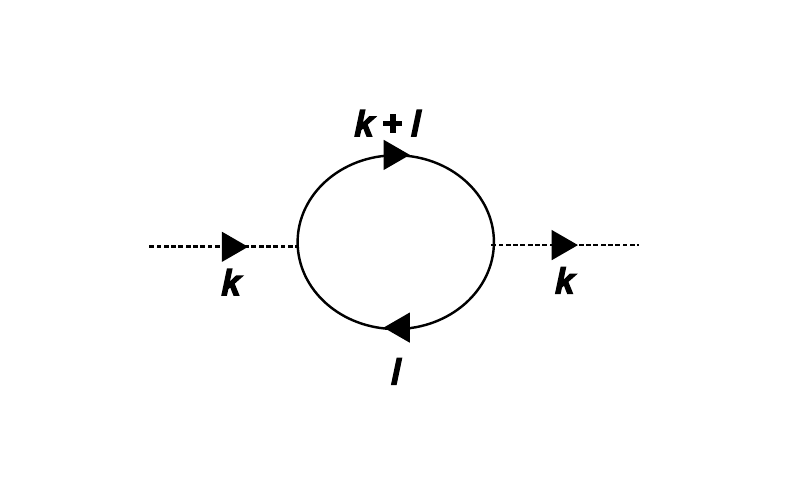}
\caption{the scalar diagram with fermion loop}
\label{scalar-fermion-loop}
\end{center}
\end{figure}

\noindent In the same way, we obtain
\begin{equation}
	\begin{split}
		&\omega_{2}((i\hbar)^{2}\,\pi_{2}\,{\bm h}\,{\bm m}_{2}\,{\bm h}\,{\bm m}_{2}\,{\bm h}\,{\bf U}\,{\bm h}\,{\bf U}\,{\bf 1},d(x_{1})\otimes d(x_{2}))\\
		&=-g^{2}\hbar^{2}\int\frac{d^{4}k}{(2\pi)^{4}}\frac{d^{4}l}{(2\pi)^{4}}\,\tilde{\Delta}(k)\,\mathrm{Tr}[\tilde{S}(l)\gamma_{5}\tilde{S}(k+l)\gamma_{5}]\,\tilde{\Delta}(k)\,e^{ik(x_{1}-x_{2})}\,. 
	\end{split}
\end{equation}
This gives the contribution of the Fourier transformation of the diagram of Figure.\ref{scalar-fermion-loop}.

\noindent The fifth term $(i\hbar)^{2}\,\pi_{2}\,{\bm h}\,{\bm m}_{2}\,{\bm h}\,{\bf U}\,{\bm h}\,{\bm m}_{2}\,{\bm h}\,{\bf U}\,{\bf 1}$ in (\ref{pi2f1-cal}) vanishes for the same reason as the one-point function of the scalar field vanishes.

Therefore, we have reproduced the one-loop correction of the two-point functions of the modified Yukawa theory using quantum $A_{\infty}$ algebras.

\section{The formula for correlation functions involving general scalar and Dirac fields}\label{Schwinger-Dyson}
\setcounter{equation}{0}
We claim that correlation functions in terms of quantum $A_{\infty}$ algebras are obtained by
\begin{equation}
\langle \, \Phi^{\otimes n} \, \rangle = \pi_n \, {\bm f} \, {\bf 1} \,,
\end{equation}
where
\begin{equation}
\Phi = \int d^d x \left[\, \varphi (x) \, c(x) + \, \overline{\theta}_\alpha (x) \, \Psi_\alpha (x)
+\overline{\Psi}_\alpha (x) \, \theta_\alpha (x) \, \right] \,,
\end{equation}
\begin{equation}
{\bm f} = \frac{1}{{\bf I} +{\bm h} \, {\bm m} +i \hbar \, {\bm h} \, {\bf U}} \,.
\end{equation}
We justify the above claim by showing that our formula satisfies the Schwinger-Dyson equations following~\cite{Okawa:2022sjf,Konosu/Okawa_2022}. In this section, we consider scalar fields and Dirac fields in general dimension $d$.

In the path integral formalism, the ordering of fields in correlation functions is (anti-)symmetric by definition. In contrast, this (anti-)symmetricity is not trivial in the homotopy algebraic context, and we need to prove the Schwinger-Dyson equations for all possible ordering of fields of the correlation functions. To handle this problem, we introduce 
\begin{equation}
	\widehat{\Phi}^i (x) = \varphi(x)\,\sigma^{\,i}+ \overline{\epsilon}^{\, i}_\alpha\,\Psi_\alpha (x)
+\overline{\Psi}_\alpha (x) \,\epsilon^i_\alpha \,,
\label{Phi-hat}
\end{equation}
where $\sigma^{\,i}$ is degree even, and $\epsilon^i_\alpha$ and $\overline{\epsilon}^{\, i}_\alpha$ are degree odd. Associated with $\widehat{\Phi}^i (x)$, we introduce
\begin{equation}
	\frac{\delta}{\delta\widehat{\Phi}^i (x)}=\sigma^{i}\,\frac{\delta}{\delta\varphi(x)}+\epsilon^{\,i}_{\alpha}\,\frac{\delta}{\delta\Psi_{\alpha}(x)}+\overline{\epsilon}^{\, i}_\alpha \, \frac{\delta}{\delta \overline{\Psi}_\alpha (x)}\,.
\end{equation}
This derivative satisfies
\begin{equation}
\frac{\delta}{\delta \widehat{\Phi}^i (x)} \, \widehat{\Phi}^j (y)
= \widehat{\delta}^{\, ij} (x-y) \,,
\end{equation}
where
\begin{equation}
\widehat{\delta}^{\, ij} (x-y)
= \delta^d (x-y) \,
( \, \sigma^{i}\,\sigma^{j}+\overline{\epsilon}^{\, i}_\alpha \, \epsilon^j_\alpha
+\overline{\epsilon}^{\, j}_\alpha \, \epsilon^i_\alpha \, ) \,.
\end{equation}
In the path integral formalism, the correlation functions for a specific ordering of fields are, for example, given by
\begin{equation}
  \begin{split}
  &\langle \, \Psi_{\alpha_1} (y_1) \, \ldots \Psi_{\alpha_m} (y_m) \,
  \overline{\Psi}_{\beta_1} (z_1) \, \ldots \overline{\Psi}_{\beta_l} (z_l) \, \varphi(x_{1})\,\ldots\,\varphi(x_{n})\,\rangle\,\\
  &=\, \frac{1}{Z}\int \mathcal{D}\varphi\,\mathcal{D}\Psi\,\mathcal{D}\overline{\Psi}\,  \Psi_{\alpha_1} (y_1) \, \ldots \Psi_{\alpha_m} (y_m) \,
  \overline{\Psi}_{\beta_1} (z_1) \, \ldots \overline{\Psi}_{\beta_l} (z_l) \,\varphi(x_{1})\,\ldots\,\varphi(x_{n}) \,e^{\frac{i}{\hbar}S},
  \end{split}
\end{equation}
where
\begin{equation}
  Z = \int \mathcal{D}\varphi\,\mathcal{D}\Psi\,\mathcal{D}\overline{\Psi}\,e^{\frac{i}{\hbar}S}.
\end{equation}
Since
\begin{equation}
  \frac{1}{Z}\int \mathcal{D}\varphi\,\mathcal{D}\Psi\,\mathcal{D}\overline{\Psi}\, \frac{\delta}{\delta\varphi(w)}  \, \Psi_{\alpha_1} (y_1) \, \ldots \Psi_{\alpha_m} (y_m) \,
  \overline{\Psi}_{\beta_1} (z_1) \, \ldots \overline{\Psi}_{\beta_l} (z_l) \,\varphi(x_{1})\,\ldots\,\varphi(x_{n})\,e^{\frac{i}{\hbar}S},\label{sd1}
\end{equation}
\begin{equation}
  \frac{1}{Z}\int \mathcal{D}\varphi\,\mathcal{D}\Psi\,\mathcal{D}\overline{\Psi}\, \frac{\delta}{\delta\overline{\Psi}_{\gamma}(w)} \, \Psi_{\alpha_1} (y_1) \, \ldots \Psi_{\alpha_m} (y_m) \,
  \overline{\Psi}_{\beta_1} (z_1) \, \ldots \overline{\Psi}_{\beta_l} (z_l) \,\varphi(x_{1})\,\ldots\,\varphi(x_{n}) \,e^{\frac{i}{\hbar}S}, \label{sd2}
\end{equation}
and
\begin{equation}
  \frac{1}{Z}\int \mathcal{D}\varphi\,\mathcal{D}\Psi\,\mathcal{D}\overline{\Psi}\, \frac{\delta}{\delta\Psi_{\gamma}(w)}  \, \Psi_{\alpha_1} (y_1) \, \ldots \Psi_{\alpha_m} (y_m) \,
  \overline{\Psi}_{\beta_1} (z_1) \, \ldots \overline{\Psi}_{\beta_l} (z_l) \,\varphi(x_{1})\,\ldots\,\varphi(x_{n}) \,e^{\frac{i}{\hbar}S}, \label{sd3}
\end{equation}
we can derive the three kinds of  Schwinger-Dyson equations. It is convenient to use $\widehat{\Phi}^i (x)$ to manipulate the above equations with arbitrary ordering of fields. Since
\begin{equation}
\frac{1}{Z} \int \mathcal{D} \varphi \, \mathcal{D} \Psi \, \mathcal{D} \overline{\Psi} \,
\frac{\delta}{\delta \widehat{\Phi}^n (x_n)} \,
\Bigl[ \, \widehat{\Phi}^1 (x_1) \, \widehat{\Phi}^2 (x_2) \, \ldots
\widehat{\Phi}^{n-1} (x_{n-1}) \, e^{\frac{i}{\hbar}S} \,
\Bigr] = 0 \,,
\end{equation}
the Schwinger-Dyson equations for all ordering of fields can be described as
\begin{equation}
\begin{split}
& \sum_{i=1}^{n-1} \, \widehat{\delta}^{\, i n} (x_i-x_n) \,
\langle \, \widehat{\Phi}^1 (x_1) \, \ldots \, \widehat{\Phi}^{i-1} (x_{i-1}) \,
\widehat{\Phi}^{i+1} (x_{i+1}) \, \ldots \, \widehat{\Phi}^{n-1} (x_{n-1}) \, \rangle \\
& \qquad +\frac{i}{\hbar} \,
\left\langle \, \widehat{\Phi}^1 (x_1) \, \ldots \, \widehat{\Phi}^{n-1} (x_{n-1}) \,
\frac{\delta S}{\delta \widehat{\Phi}^n (x_n)} \, \right\rangle = 0 \,.
\end{split}
\label{Dirac-Schwinger-Dyson}
\end{equation}

Let us confirm that correlation functions described in terms of quantum $A_{\infty}$ algebras satisfy the Schwinger-Dyson equations. The proof of the free part is essentially the same as in~\cite{Konosu/Okawa_2022}, so in this paper, we mainly discuss how to deal with interaction terms.  The key identity for the proof is
\begin{equation}
	(\,{\bf I}+{\bm h}{\bm m}+i\hbar\,{\bm h}{\bf U}\,)\,{\bm f}={\bf I}\,.
\end{equation}
Then, we act the projection $\pi_{n}$ from the left-hand side and ${\bf 1}$ from the right-hand side to obtain
\begin{equation}
	\pi_{n}\,{\bm f}\,{\bf 1}+\pi_{n}\,{\bm h}\,{\bm m}\,{\bm f}\,{\bf 1}+i\hbar\,\pi_{n}\,{\bm h}\,{\bf U}\,{\bm f}\,{\bf 1}=0\,.\label{main-homotopy-SD}
\end{equation}
for $n\geq1$.
To prove the extended version of the Schwinger-Dyson equations~\eqref{Dirac-Schwinger-Dyson} directly, we introduce $\widehat{\lambda}^{\, i} (x)$ by
\begin{equation}
\widehat{\lambda}^{\, i} (x) = d(x)\,\sigma^{\,i}-\overline{\epsilon}^{\, i}_\alpha \, \lambda_\alpha (x)
-\overline{\lambda}_\alpha (x) \, \epsilon^i_\alpha \,.
\end{equation}
and this satisfies
\begin{equation}
	\begin{split}
		&\omega\,(\,c(x')\,,\, \widehat{\lambda}^{\, i} (x) \, ) =\sigma^{i}\,\delta(x'-x)\,,\\
		&\omega \, ( \, \overline{\theta}_{\alpha'} (x') \,,\, \widehat{\lambda}^{\, i} (x) \, ) 
			= {}-\overline{\epsilon}^{\, i}_{\alpha'} \, \delta^d ( x'-x ) \,, \\
		&\omega \, ( \, \theta_{\alpha'} (x') \,,\, \widehat{\lambda}^{\, i} (x) \, )
		= {}-\epsilon^i_{\alpha'} \, \delta^d ( x'-x ) \,.
	\end{split}
\end{equation}
This extended basis vector $\widehat{\lambda}^{\, i} (x)$ enables us to extend our formula:
\begin{equation}
\langle \, \widehat{\Phi}^1 (x_1) \, \widehat{\Phi}^2 (x_2) \, \ldots \,
\widehat{\Phi}^n (x_n) \, \rangle
=\omega_n \, ( \, \pi_n \, {\bm f} \, {\bf 1} \,,\,
\widehat{\lambda}^{\, 1} (x_1) \otimes \widehat{\lambda}^{\, 2} (x_2) \otimes \ldots
\otimes \widehat{\lambda}^{\, n} (x_n) \, ) \,.
\label{extended-formula}
\end{equation}
The extended basis vector $\widehat{\lambda}^{\, i} (x)$ enables us to deduce the relation from~\eqref{main-homotopy-SD}: 
\begin{equation}
	\begin{split}
		&\omega_{n}(\,\pi_{n}\,{\bm f}\,{\bf 1}\,,\,\widehat{\lambda}^{\, 1} (x_1) \otimes \widehat{\lambda}^{\, 2} (x_2) \otimes \ldots \otimes \widehat{\lambda}^{\, n} (x_n) \,)\\
		&\quad+i\hbar\, \omega_{n}(\,\pi_{n}\,{\bm h}\,{\bf U}\,{\bm f}\,{\bf 1}\,,\,\widehat{\lambda}^{\, 1} (x_1) \otimes \widehat{\lambda}^{\, 2} (x_2) \otimes \ldots \otimes \widehat{\lambda}^{\, n} (x_n) \,)\\
		&\quad+\omega_{n}(\,\pi_{n}\,{\bm h}\,{\bm m}\,{\bm f}\,{\bf 1}\,,\,\widehat{\lambda}^{\, 1} (x_1) \otimes \widehat{\lambda}^{\, 2} (x_2) \otimes \ldots \otimes \widehat{\lambda}^{\, n} (x_n) \,)=0\,.\label{Dirac-omega_n}
	\end{split}
\end{equation}
Let us consider each term on the left-hand side.
As in the same discussion in the previous paper, the first term on the left-hand side corresponds to the $n$-point function:
\begin{equation}
	\omega_n \, ( \, \pi_n \, {\bm f} \, {\bf 1} \,,\,
	\widehat{\lambda}^{\, 1} (x_1)  \otimes \widehat{\lambda}^{\, 2} (x_2)  \otimes \ldots
	\otimes \widehat{\lambda}^{\, n} (x_n) \, )
	= \langle \, \widehat{\Phi}^1 (x_1) \, \widehat{\Phi}^2 (x_2) \, \ldots \,
	\widehat{\Phi}^n (x_n) \, \rangle \,.
\end{equation}
In the same way as the previous paper, the second term on the left-hand side of~\eqref{Dirac-omega_n} is
\begin{equation}
\begin{split}
& i \hbar \, \omega_n \, ( \, \pi_n \, {\bm h} \, {\bf U} \, {\bm f} \, {\bf 1} \,,\,
\widehat{\lambda}^{\, 1} (x_1)  \otimes \widehat{\lambda}^{\, 2} (x_2)  \otimes \ldots
\otimes \widehat{\lambda}^{\, n} (x_n) \, ) \\
& = {}-\frac{\hbar}{i} \, \sum_{i=1}^{n-1} \widehat{S}^{\, in} (x_i-x_n) \,
\langle\, \widehat{\Phi}^1 (x_1) \, \ldots \,
\widehat{\Phi}^{i-1} (x_{i-1}) \,
\widehat{\Phi}^{i+1} (x_{i+1}) \, \ldots \,
\widehat{\Phi}^{n-1} (x_{n-1}) \, \rangle 
\end{split}
\end{equation}
for $n>1$, where
\begin{equation}
\widehat{S}^{\, ij} (x_i-x_j)
= \sigma^{i}\Delta(x_{i}-x_{j})\,\sigma^{j}+\overline{\epsilon}^{\, i}_\alpha \, S (x_i-x_j)_{\alpha \beta} \, \epsilon^j_\beta
+\overline{\epsilon}^{\, j}_\alpha \, S (x_j-x_i)_{\alpha \beta} \, \epsilon^i_\beta\,.
\end{equation}
Let us consider the third term on the  left-hand side of~\eqref{Dirac-omega_n}
\begin{equation}
	\omega_{n}(\,\pi_{n}\,{\bm h}\,{\bm m}\,{\bm f}\,{\bf 1}\,,\,\widehat{\lambda}^{\, 1} (x_1) \otimes \widehat{\lambda}^{\, 2} (x_2) \otimes \ldots \otimes \widehat{\lambda}^{\, n} (x_n) \,)\,.\label{third-SD}
\end{equation}
Since
\begin{equation}
	\begin{split}
		\delta_{\varphi}S&=-\sum_{n=1}^{\infty}\,\omega(\,\delta_{\varphi}\Phi,\,M_{n}(\Phi\otimes\ldots\otimes\Phi)\,)\\
				              &=-\sum_{n=1}^{\infty}\,\int d^{d}x \,\delta\varphi(x)\,\omega(c(x),\,M_{n}(\Phi\otimes\ldots\otimes\Phi))\\
				              &=\int d^{d}x\,\delta\varphi(x)\,\frac{\delta S}{\delta \varphi(x)}\,,
	\end{split}
\end{equation}
\begin{equation}
	\begin{split}
		\delta_{\Psi}S&=-\sum_{n=1}^{\infty}\,\omega(\,\delta_{\Psi}\Phi,\,M_{n}(\Phi\otimes\ldots\otimes\Phi)\,)\\
				              &=-\sum_{n=1}^{\infty}\,\int d^{d}x \,\delta\Psi_{\alpha}(x)\,\omega(\theta_{\alpha}(x),\,M_{n}(\Phi\otimes\ldots\otimes\Phi))\\
				              &=\int d^{d}x\,\delta\Psi_{\alpha}(x)\,\frac{\delta S}{\delta \Psi_{\alpha}(x)}\,,
	\end{split}
\end{equation}
and
\begin{equation}
	\begin{split}
		\delta_{\overline{\Psi}}S&=-\sum_{n=1}^{\infty}\,\omega(\,\delta_{\overline{\Psi}}\Phi,\,M_{n}(\Phi\otimes\ldots\otimes\Phi)\,)\\
				              &=\sum_{n=1}^{\infty}\,\int d^{d}x \,\delta\overline{\Psi}(x)\,\omega(\theta_{\alpha}(x),\,M_{n}(\Phi\otimes\ldots\otimes\Phi))\\
				              &=\int d^{d}x\,\delta\overline{\Psi}_{\alpha}(x)\,\frac{\delta S}{\delta \overline{\Psi}_{\alpha}(x)}\,,
	\end{split}
\end{equation}
we find\footnote{Thank you to Jojiro Totsuka-Yoshinaka for telling me this relation.}
\begin{equation}
	\sum_{n=0}^{\infty}\,M_{n}(\Phi\otimes\ldots\otimes\Phi)={}-\int\,d^{d}x\,\left[d(x)\frac{\delta S}{\delta \varphi(x)}+\lambda_{\alpha}(x)\frac{\delta S}{\delta\Psi_{\alpha}(x)}-\overline{\lambda}_{\alpha}(x)\frac{\delta S}{\delta\overline{\Psi}_{\beta}(x)}\right]\,.
\end{equation}
This relation enable us to transform the simple form of $\pi_{n}\,{\bm h}\,{\bm m}\,{\bm f}\,{\bf 1}$:
\begin{equation}
	\begin{split}
		\pi_{n}\,{\bm h}\,{\bm m}\,{\bm f}\,{\bf 1}&=\pi_{n}\sum_{k=0}^{\infty}\,{\bm h}\,{\bm m}_{k}\,\langle\,\Phi^{\otimes (n+k-1)}\,\rangle\\
							            &=\sum_{k=0}^{\infty}\langle\,\Phi^{\otimes (n-1)}\otimes hm_{k}(\Phi\otimes\ldots\Phi)\,\rangle\\
							            &=-\int d^{d}y\,\langle\,\Phi^{\otimes(n-1)}\otimes\left[hd(y)\frac{\delta S_{\mathrm{int}}}{\delta \varphi(y)}+h\lambda_{\alpha}(y)\frac{\delta S_{\mathrm{int}}}{\delta\Psi_{\alpha}(y)}-h\overline{\lambda}_{\alpha}(y)\frac{\delta S_{\mathrm{int}}}{\delta\overline{\Psi}_{\beta}(y)}\right]\,\rangle\\
							            &=-\int d^{d}x\,d^{d}y\,\Big\langle\,\Phi^{\otimes(n-1)}\\
							            &\otimes\left[\Delta(y-x)c(x)\frac{\delta S_{\mathrm{int}}}{\delta \varphi(y)}+S(y-x)_{\alpha\beta}\theta_{\beta}(x)\frac{\delta S_{\mathrm{int}}}{\delta\Psi_{\alpha}(y)}+\overline{\theta}_{\beta}(x)S(y-x)_{\beta\alpha}\frac{\delta S_{\mathrm{int}}}{\delta\overline{\Psi}_{\beta}(y)}\right]\,\Big\rangle\,,
	\end{split}
\end{equation}
where $S_{\mathrm{int}}$ is the interaction part of the action $S$. Therefore, the term~\eqref{third-SD} can be simplified as
\begin{equation}
	\begin{split}
		& \omega_{n}(\,\pi_{n}\,{\bm h}\,{\bm m}\,{\bm f}\,{\bf 1}\,,\,\widehat{\lambda}^{\, 1} (x_1) \otimes \widehat{\lambda}^{\, 2} (x_2) \otimes \ldots \otimes \widehat{\lambda}^{\, n} (x_n) \,)\\
		&= {}-\int d^{d}y\,\langle\,\widehat{\Phi}^1 (x_1) \, \widehat{\Phi}^2 (x_2) \, \ldots \,\widehat{\Phi}^{n-1} (x_{n-1})[\Delta(y-x_{n})\sigma^{n}+S(x_{n}-y)_{\alpha_{n}\alpha}\overline{\epsilon}^{n}_{\alpha_{n}}\frac{\delta S_{\mathrm{int}}}{\delta\overline{\Psi}_{\alpha}(y)}\\
		&\qquad +S(y-x_{n})_{\alpha\alpha_{n}}\epsilon^{n}_{\alpha_{n}}\frac{\delta S_{\mathrm{int}}}{\delta\Psi_{\alpha}(y)}]\,\rangle\,.	
	\end{split}
\end{equation}
To summarize the above results, the relation~\eqref{Dirac-omega_n} for $n>1$ can be translated into
\begin{equation}
	\begin{split}
		&\langle \, \widehat{\Phi}^1 (x_1) \, \widehat{\Phi}^2 (x_2) \, \ldots \,\widehat{\Phi}^n (x_n)\,\rangle\\
		&\quad-\frac{\hbar}{i} \, \sum_{i=1}^{n-1} \widehat{S}^{\, in} (x_i-x_n) \,\langle\, \widehat{\Phi}^1 (x_1) \, \ldots \,\widehat{\Phi}^{i-1} (x_{i-1}) \,\widehat{\Phi}^{i+1} (x_{i+1}) \, \ldots \,\widehat{\Phi}^{n-1} (x_{n-1}) \, \rangle\\
		& {}-\int d^{d}y\,\Big\langle\,\widehat{\Phi}^1 (x_1) \, \widehat{\Phi}^2 (x_2) \, \ldots \,\widehat{\Phi}^{n-1} (x_{n-1})[\Delta(y-x_{n})\sigma^{n}+S(x_{n}-y)_{\alpha_{n}\alpha}\overline{\epsilon}^{n}_{\alpha_{n}}\frac{\delta S_{\mathrm{int}}}{\delta\overline{\Psi}_{\alpha}(y)}\\
		&\qquad +S(y-x_{n})_{\alpha\alpha_{n}}\epsilon^{n}_{\alpha_{n}}\frac{\delta S_{\mathrm{int}}}{\delta\Psi_{\alpha}(y)}]\,\Big\rangle\\
		&\quad=0\,. \label{SD-before}
	\end{split}
\end{equation}
Then, we act the operator
\begin{equation}
\widehat{D}^{n}_{x_{n}}=\sigma^{n}(-\partial_{x_{n}}^{2}+M^{2})\frac{\partial}{\partial\sigma^{n}}+\epsilon^n_\alpha \, ( \, i \, \partial\!\!\!/_{x_n} +m \, )_{\beta \alpha} \,
\frac{\partial}{\partial \epsilon^n_\beta}
+\overline{\epsilon}^{\, n}_\alpha \, ( \, -i \, \partial\!\!\!/_{x_n} +m \, )_{\alpha \beta} \,
\frac{\partial}{\partial \overline{\epsilon}^{\, n}_\beta}
\label{Q-hat}
\end{equation}
on~\eqref{SD-before}. Notice that
\begin{equation}
		\widehat{D}^{n}_{x_{n}}\,\widehat{S}^{\, in} (x_i-x_n)= \widehat{\delta}^{\, in} (x_i-x_n) \,,
\end{equation}
and
\begin{equation}
\begin{split}
\widehat{D}^{n}_{x_{n}}
\widehat{\Phi}^n (x_n) = (-\partial_{x_{n}}^{2}+M^{2})\,\varphi(x_{n})\,\sigma^{n}+\overline{\Psi}_\beta (x_n) \,
( \, i \overleftarrow{\partial\!\!\!/}\!_{x_n} +m \, )_{\beta \alpha} \, \epsilon^n_\alpha
+\overline{\epsilon}^{\, n}_\alpha
( \, -i \, \partial\!\!\!/_{x_n} +m \, )_{\alpha \beta} \, \Psi_\beta (x_n) \,.
\end{split}
\end{equation}
Since
\begin{equation}
\frac{\delta S_{\mathrm{free}}}{\delta \widehat{\Phi}^n (x_n)}
= (\partial^{2}_{x_{n}}-M^{2})\,\varphi(x_{n})-\overline{\Psi}_\beta (x_n) \,
( \, i \overleftarrow{\partial\!\!\!/}\!_{x_n} +m \, )_{\beta \alpha} \, \epsilon^n_\alpha
+\overline{\epsilon}^{\, n}_\alpha
( \, i \, \partial\!\!\!/_{x_n} -m \, )_{\alpha \beta} \, \Psi_\beta (x_n)\,,
\end{equation}
we obtain
\begin{equation}
		 \widehat{D}^{n}_{x_{n}}\widehat{\Phi}^n (x_n)= {}-\frac{\delta S_{\mathrm{free}}}{\delta \widehat{\Phi}^n (x_n)} \,.
\end{equation}
Moreover,
\begin{equation}
	\begin{split}
		&\widehat{D}^{n}_{x_{n}}\{-\int d^{d}y [\Delta(y-x_{n})\sigma^{n}+S(x_{n}-y)_{\alpha_{n}\alpha}\overline{\epsilon}^{n}_{\alpha_{n}}\frac{\delta S_{\mathrm{int}}}{\delta\overline{\Psi}_{\alpha}(y)} +S(y-x_{n})_{\alpha\alpha_{n}}\epsilon^{n}_{\alpha_{n}}\frac{\delta S_{\mathrm{int}}}{\delta\Psi_{\alpha}(y)}]\}\\
		&={}-\frac{\delta S_{\mathrm{int}}}{\delta \widehat{\Phi}(x_{n})}\,.
	\end{split}
\end{equation}
Then, the action of the operator $\widehat{D}^{n}_{x_{n}}$ on~\eqref{SD-before} yields
\begin{equation}
	\begin{split}
		&-\langle \, \widehat{\Phi}^1 (x_1) \, \widehat{\Phi}^2 (x_2) \, \ldots \,\widehat{\Phi}^{n-1} (x_{n-1})\,\frac{\delta S}{\delta\widehat{\Phi}(x_{n})}\,\rangle\\
		&\quad-\frac{\hbar}{i} \, \sum_{i=1}^{n-1} \widehat{\delta}^{\, in} (x_i-x_n) \,\langle\, \widehat{\Phi}^1 (x_1) \, \ldots \,\widehat{\Phi}^{i-1} (x_{i-1}) \,\widehat{\Phi}^{i+1} (x_{i+1}) \, \ldots \,\widehat{\Phi}^{n-1} (x_{n-1}) \, \rangle\\
		&\quad=0
	\end{split}
\end{equation}
for $n>1$.
This is exactly the Schwinger-Dyson equation~\eqref{Dirac-Schwinger-Dyson} for $n>1$. For $n=1$, the second term on the left-hand side is absent since
\begin{equation}
	\pi_1 \, {\bm h} \, {\bf U}  = 0
\end{equation}
and this reproduces the Schwinger-Dyson equations for $n=1$.

\section{Conclusions and Discussions}\label{conclusion-section}
\setcounter{equation}{0}
In this paper, we extend the result~\cite{Konosu/Okawa_2022} to  general scalar-Dirac systems. The formula presented in the previous paper is given by 
\begin{equation}
\langle \, \Phi^{\otimes n} \, \rangle = \pi_n \, {\bm f} \, {\bf 1} \,,
\label{formula-conclusions}
\end{equation}
where
\begin{equation}
{\bm f} = \frac{1}{{\bf I} +{\bm h} \, {\bm m} +i \hbar \, {\bm h} \, {\bf U}} \,,
\end{equation}
and
\begin{equation}
\Phi = \int d^d x \, \varphi (x) \, c(x)
+\int d^d x \, ( \, \overline{\theta}_\alpha (x) \, \Psi_\alpha (x)
+\overline{\Psi}_\alpha (x) \, \theta_\alpha (x) \, ) \,.
\end{equation}
Since we dealt with free theory in the previous paper, the operator $ {\bm m}$ vanishes:
\begin{equation}
	 {\bm m}=0\,.
\end{equation}
In this paper, we have shown that the formula~\eqref{formula-conclusions} holds in  general scalar-Dirac systems, which means
\begin{equation}
	 {\bm m}\neq0\,.
\end{equation}
We can extend this formalism to the other types of fermions such as Majorana fields.

As we discussed in the previous paper, in spite of the asymmetric construction of ${\bm h}$,
$\pi_n \, {\bm f} \, {\bf 1}$ is
totally symmetric for bosonic fields
and totally antisymmetric for fermionic fields. Surprisingly,  this property holds for the interacting theories. This should be directly proved using the techniques of coalgebra representations of $A_{\infty}$ algebras and we will try to prove this in future work.

Most of the discussions and future directions are discussed in the previous paper, so see~\cite{Konosu/Okawa_2022} in detailed discussions. 
We briefly state them. 

One future direction is a mathematical one. We believe that it is important to discuss the
relation to the approach of factorization algebras developed by Costello and Gwilliam~\cite{Costello:2016vjw, Costello:2021jvx}. This would contribute to revealing the mathematical nature of quantum field theory. 

The other direction is to generalize our approach based on $A_\infty$ algebras to open superstring field theory. For instance, it is discussed in~\cite{Okawa:2020llq} that the $1/N$ expansion of correlation functions of gauge-invariant operators in open superstring field theory is important in the program of providing a framework to prove the AdS/CFT correspondence, but correlation functions have not been discussed much in string field theory.\footnote{
See recent interesting progress~\cite{Maccaferri:2023gcg, Maccaferri:2023gof}
which will be relevant to this topic.
It would also be intriguing to consider
the twisted holography~\cite{Costello:2016mgj, Costello:2017fbo, Costello:2018zrm} in our context,
and see~\cite{Zeng:2023qqp} for recent research.
}

We hope that this work will contribute to the understanding of the mathematical nature of quantum field theory and the quantization of open string field theory.

\bigskip
\noindent
{\normalfont \bfseries \large Acknowledgments}

We wish to thank  Yuji Okawa for  discussions and comments on drafts.

\appendix
\setcounter{equation}{0}
\section{Yukawa theory}\label{Yukawa}
\setcounter{equation}{0}
In this paper, we use string-field-theory-like expressions to express quantum field theory in terms of $A_{\infty}$ algebras. In this formalism, we will calculate the tree-level scattering amplitudes of the Yukawa theory\footnote{Calculations of the tree-level amplitudes of Yukawa theory in the path integral formalism are presented, for example, in section 45 of~\cite{Srednicki:2007qs}.} using $A_{\infty}$ algebras.

First, we denote some notations. The free Dirac fields in $d=4$ dimensions can be expanded as
\begin{equation}
	\Psi(x)=\sum_{s=\pm}\int\widetilde{dp} \,\left[\,b_{s}({\bf p})\,u_{s}({\bf p})\,e^{ipx}+d^{\dagger}_{s}({\bf p})\,v_{s}({\bf p})\,e^{-ipx}\,\right]\,,
\end{equation}
where
\begin{equation}
	\int\widetilde{dp}=\frac{d^{3}p}{(2\pi)^{3}2\omega}\,,
\end{equation}
\begin{equation}
	\omega=p^{0}\,,
\end{equation}
$b_{s}({\bf p})$ and $d_{s}({\bf p})$ are Grassmann number, and $u_{s}({\bf p})$ and $v_{s}({\bf p})$ are four-component vectors. The complex numbers $b_{s}({\bf p}),\,d_{s}({\bf p})$ correspond to annihilation operators of electrons and positrons, respectively, after the canonical quantization.

\subsection{Action}
We consider an action
\begin{equation}
  S = \int \,d^4 x\, \left[-\frac{1}{2}\,\partial^{\mu}\varphi(x) \partial_{\mu}\varphi(x) \,-\frac{1}{2}\,M^2\varphi^2\,+i\overline{\Psi}(x){}\partial\!\!\!/\Psi(x)-m\overline{\Psi}(x)\Psi(x)\,+g\varphi(x)\overline{\Psi}(x)\Psi(x)\right].
\end{equation}
Since we only consider tree-level amplitudes, we omit the renormalization parameters.
Again, we consider the vector space
\begin{equation}
	\mathcal{H}=\mathcal{H}_{1}\oplus\mathcal{H}_{2}\,,
\end{equation}
and we rewrite this action to the form 
\begin{equation}
S = {}-\frac{1}{2} \, \omega \, ( \, \Phi, Q \, \Phi \, )
-\, \frac{1}{3} \,
\omega \, ( \, \Phi \,, m_2 \, ( \, \Phi \otimes \Phi \, ) \, ) \,,
\end{equation}
where $\Phi$ in $\mathcal{H}_1$ is expanded as
\begin{equation}
\Phi = \int d^4 x \left[\, \varphi (x) \, c(x) + \, \overline{\theta}_\alpha (x) \, \Psi_\alpha (x)
+\overline{\Psi}_\alpha (x) \, \theta_\alpha (x) \, \right] \,.
\end{equation}
The definitions of $\omega,\,Q$ are the same as in the previous sections. 
We define $m_2$ as
\begin{equation}
  \begin{split}
    m_2(\theta_{\alpha}(x_1)\,\otimes\,\overline{\theta}_{\beta}(x_2))\,&=\,-\frac{g}{2}\,\delta_{\alpha \beta}\,\delta^4(x_1-x_2)\,d(x_1), \\
    m_2(\overline{\theta}_{\alpha}(x_1)\,\otimes\,\theta_{\beta}(x_2))\,&=\,\,\,\,\,\,\frac{g}{2}\,\delta_{\alpha \beta}\,\delta^4(x_1-x_2)\,d(x_1), \\
    m_2(\,c\,(x_1)\,\otimes\,\theta_{\alpha}(x_2))\,\,\,&=\,-\frac{g}{2}\,\delta^4(x_1-x_2)\,\lambda_{\alpha}(x_1), \\
    m_2(\theta_{\alpha}(x_1)\otimes\,\,c\,(x_2)\,)\,\,\,&=\,-\frac{g}{2}\,\delta^4(x_1-x_2)\,\lambda_{\alpha}(x_1), \\
    m_2(\overline{\theta}_{\alpha}(x_1)\otimes\,\,c\,(x_2)\,)\,\,\,&=\,\,\,\,\,\,\frac{g}{2}\,\delta^4(x_1-x_2)\,\overline{\lambda}_{\alpha}(x_1), \\
    m_2(\,c\,(x_1)\,\otimes\,\overline{\theta}_{\alpha}(x_2))\,\,\,&=\,\,\,\,\,\,\frac{g}{2}\,\delta^4(x_1-x_2)\,\overline{\lambda}_{\alpha}(x_1)
  \end{split}
\end{equation}
to obtain
\begin{equation}
  -\frac{1}{3}\,\omega(\,\Phi,\,m_2(\Phi\,\otimes\,\Phi))= g\,\int d^4 x \,\varphi(x)\,\overline{\Psi}(x)\,\Psi(x)\,.
\end{equation}
We thus obtain the action of the Yukawa theory in terms of quantum $A_{\infty}$ algebras.

\subsection{Tree-level scattering amplitudes}
It is known that the $n$-point scattering amplitude $\mathcal{A}$ is calculated in terms of $A_{\infty}$ algebras by the following formula:\footnote{See for example~\cite{Kajiura:2003ax,Macrelli:2019afx}. See also~\cite{Macrelli:2021apq} for the comprehensive review.}
\begin{equation}
  \mathcal{A} = -i\sum_{\sigma\in S_{n-1}}\omega(\phi_{1},\widetilde{m}_{n-1}(\phi_{\sigma(2)},\phi_{\sigma(3)},\ldots,\phi_{\sigma(n)}))\,,
\end{equation}
where  
\begin{equation}
    \widetilde{m}_{n}\,\pi_{n} = \pi_{1} \,{\bf P}\,{\bm m}\,{\bm f}^{(0)}\,{\bf P}\,\pi_{n}\,, \label{minimal}
\end{equation}
\begin{equation}
  \bm{f}^{(0)}=\frac{1}{{\bf I}\,+\,\bm{h}\,\bm{m}}\,,
\end{equation}
$S_{n}$ is the n-th order symmetry group, and $\sigma$ is its permutation.
In this case, we need to consider different contracting homotopy $h$ from the previous sections. To calculate scattering amplitudes, we consider the on-shell projector $P$. Using this projector, we define
\begin{equation}
\begin{split}
h \, c(x) = 0 \,, \qquad
&h \, d(x) = \int d^4 y \, \Delta (x-y) \,(\,\mathbb{I}-P\,)\, c (y) \,,\\
h \, \theta_\alpha (x) = 0 \,, \qquad
&h \, \lambda_\alpha (x) = {}\int d^4 y \, S(x-y)_{\alpha \beta}  \,(\,\mathbb{I}-P\,)\, \theta_\beta (y) \,, \\
h \, \overline{\theta}_\alpha (x) = 0 \,, \qquad
&h \, \overline{\lambda}_\alpha (x) = -\int d^4 y \,(\,\mathbb{I}-P\,)\, \overline{\theta}_\beta (y) \, S(y-x)_{\beta \alpha}  \,.
\end{split}
\end{equation}
These definitions satisfy the Hodge-Kodaira decomposition and the annihilation conditions
\begin{equation}
Q \, h +h \, Q = \mathbb{I}-P \,, \quad
h \, P = 0 \,, \quad
P \, h = 0 \,, \quad
h^2 = 0 \,.
\end{equation}
Then, we have defined all ingredients to calculate the amplitudes. 

\begin{figure}[htbp]    
\begin{center}      
\includegraphics[width=8 cm]{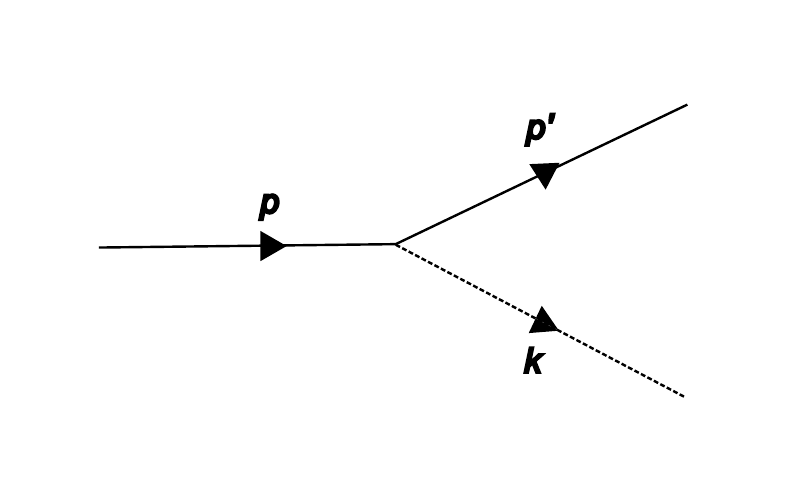}
\caption{$e^{-}\rightarrow e^{-}\varphi$}
\label{3pt-amp}
\end{center}
\end{figure}

First, we consider 3-point amplitude $e^{-}\rightarrow e^{-}\varphi$. This amplitude can be calculated by considering
\begin{equation}
  \mathcal{A}_{3} = -i[\,\omega(\phi,\widetilde{m}_{2}(\psi_{1},\overline{\psi}_{2}))+\omega(\phi,\widetilde{m}_{2}(\overline{\psi}_{2},\psi_{1}))]\,. \label{3pt}
\end{equation}
We take $\phi$, $\psi_{1}$, $\overline{\psi}_{2}$ to be
\begin {align}
        \phi &= \int d^{4}x_{1}\,e^{-ikx_{1}}\,c(x_{1})\,, \label{3phi}\\ 
        \psi_{1} &=  \int d^{4}x_{2}\,\overline{\theta}_{\alpha}(x_{2})\,u_{s}({\bf p})_{\alpha}\,\xi\, e^{ipx_{2}}\,,\\
        \overline{\psi}_{2} &=\int d^{4}x_{3}\,\overline{u}_{s'}({\bf p}')_{\beta}\,e^{-ip'x_{3}}\,\overline{\eta}\,\theta_{\beta}(x_{3})\,,
    \end {align}
where we introduced degree-odd parameters $\xi$ and $\overline{\eta}$. Note that the amplitudes become the coefficient of degree-odd parameters. In this paper, we do not distinguish amplitudes multiplied by degree-odd parameters and true amplitudes.\footnote{In general, amplitudes contains sign ambiguity. The motivation  we introduce the degree-odd parameters is to relate the Lehmann-Symanzik-Zimmermann (LSZ) reduction formula and homotopy algebraic calculations. We conjecture that the ordering of degree-odd parameters corresponds to the ordering of the creation-annihilation operators in the LSZ reduction formula.} 
From the definition (\ref{minimal}), we obtain
\begin{equation}
  \begin{split}
     \widetilde{m}_{2} &= \pi_{1} {\bf P}\,{\bm m}_{2}\,\frac{1}{{\bf I}\,+\,\bm{h}\,\bm{m}_{2}}\,{\bf P}\,\pi_{2}\\
     			&= P\,m_{2}\,(P\otimes P)\,\pi_{2}\,. \label{2-string-op}
  \end{split}
\end{equation}
Let us calculate the first term in (\ref{3pt}). From the result (\ref{2-string-op}), we obtain
\begin{equation}
 	 \widetilde{m}_{2} (\,\psi_{1}\,,\,\overline{\psi}_{2}\,)=\int d^{4}\,x_{2}\,d^{4}\,x_{3}\,\overline{u}_{s'}({\bf p}')_{\beta}\,u_{s}({\bf p})_{\alpha}\,e^{-ip'x_{3}}\,e^{ipx_{2}}\,m_{2}(\,\overline{\theta}_{\alpha}(x_{2})\,\xi\,,\,\overline{\eta}\,\theta_{\beta}(x_{3})\,)\,.
\end{equation}
We need to care about the sign factors when we calculate $m_{2}(\,\overline{\theta}_{\alpha}(x_{2})\,\xi\,,\,\overline{\eta}\,\theta_{\beta}(x_{3})\,)$.
\begin{equation}
	\begin{split}
		m_{2}(\,\overline{\theta}_{\alpha}(x_{2})\,\xi\,,\,\overline{\eta}\,\theta_{\beta}(x_{3})\,)&=\xi\,m_{2}\,(\overline{\theta}_{\alpha}(x_{2}),\overline{\eta}\,\theta_{\beta}(x_{3}))\\
																	      &={}-\overline{\eta}\,\xi\,m_{2}(\overline{\theta}_{\alpha}(x_{2}),\theta_{\beta}(x_{3}))
	\end{split}
\end{equation}
Then, we obtain
\begin{equation}
 	 \widetilde{m}_{2} (\,\psi_{1}\,,\,\overline{\psi}_{2}\,)={}-\frac{g}{2}\int d^{4}\,x_{2}\,\overline{u}_{s'}({\bf p}')\,u_{s}({\bf p})\,e^{-i(p'-p)x_{2}}\,d(x_{2})\,\overline{\eta}\,\xi\,.\label{3m2}
\end{equation}
We can calculate $\widetilde{m}_{2} (\,\psi_{1}\,,\,\overline{\psi}_{2}\,)$ in the same way, and we find
\begin{equation}
 	  \widetilde{m}_{2} (\,\overline{\psi}_{2}\,,\,\psi_{1}\,)=\widetilde{m}_{2} (\,\psi_{1}\,,\,\overline{\psi}_{2}\,)\,.
\end{equation}
Then, we obtain
\begin{equation}
  \mathcal{A}_{3} = {}-2i\,\omega(\phi,\tilde{m}_{2}(\,\psi_{1},\,\overline{\psi}_{2}))\,.
\end{equation}
Finally, if we substitute (\ref{3phi}) and (\ref{3m2}) into the above, we obtain
\begin{equation}
  \mathcal{A}_{3} = {}(2\pi)^{4}\,i\,g\,\delta^{4}(p\,-\,(k\,+\,p'))\,\overline{u}_{s'}({\bf p}')\,u_{s}({\bf p})\,\overline{\eta}\,\xi\,.
\end{equation}
This completely agrees with the result using the Feynman diagram (Figure \ref{3pt-amp}).

Next, let us calculate 4-point amplitudes $\mathcal{A}_{4}$. The amplitudes $\mathcal{A}_{4}$ can be calculated from following formula:
\begin{equation}
	\begin{split}
    		\mathcal{A}_{4} = -i&[\,\omega(\phi_{1},\widetilde{m}_{3}(\phi_{2},\phi_{3},\phi_{4}))+\omega(\phi_{1},\widetilde{m}_{3}(\phi_{2},\phi_{4},\phi_{3}))\\
									&+\omega(\phi_{1},\widetilde{m}_{3}(\phi_{3},\phi_{4},\phi_{2}))+\omega(\phi_{1},\widetilde{m}_{3}(\phi_{3},\phi_{2},\phi_{4}))\\
									&+\omega(\phi_{1},\widetilde{m}_{3}(\phi_{4},\phi_{3},\phi_{2}))+\omega(\phi_{1},\widetilde{m}_{3}(\phi_{4},\phi_{2},\phi_{3}))]\,.
	\end{split}
\end{equation}
From the definition (\ref{minimal}), we can calculate $\widetilde{m}_{3}$ as follows:
\begin{equation}
	\begin{split}
		\widetilde{m}_{3}\,\pi_{3} & =\pi_1 \,{\bf{P}}\, {\bm{m}}_2 \,\frac{1}{{\bf{I}}+{\bm h}\,{\bm{m}}_2} \,{\bf{P}}\, \pi_3 \\
				& =-\pi_1\, {\bf{P}} \,{\bm{m}}_2 \,{\bm{h}} \,{\bm{m}}_2 \,{\bf{P}}\, \pi_3 \\
				& =-P\, m_2\left(h \,m_2 \otimes P+I \otimes h\, m_2\right)(P \otimes P \otimes P)\,\pi_{3}\,.
	\end{split}
\end{equation}

\begin{figure}[htbp]
  \begin{minipage}[b]{0.45\linewidth}
 \begin{center}      
\includegraphics[width =7.5 cm]{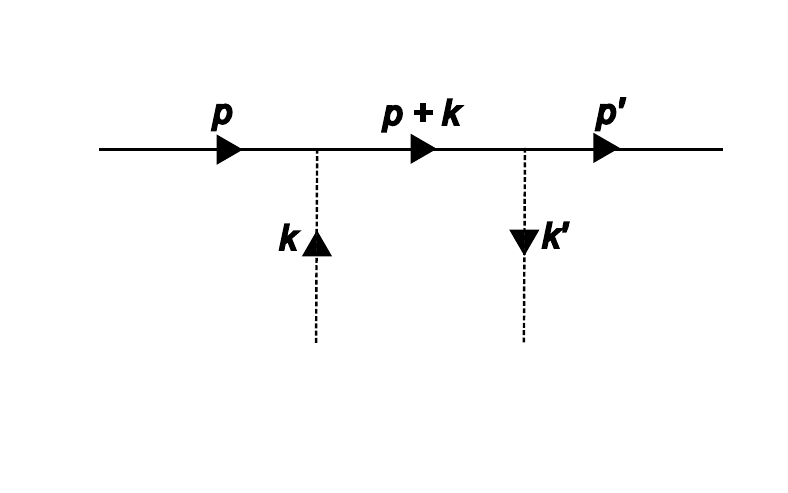}
\caption{$e^{-}\varphi\rightarrow e^{-}\varphi$}
\label{4-scattering-1}
\end{center}
  \end{minipage}
  \begin{minipage}[b]{0.45\linewidth}
   \begin{center}     
\includegraphics[width =7.5 cm]{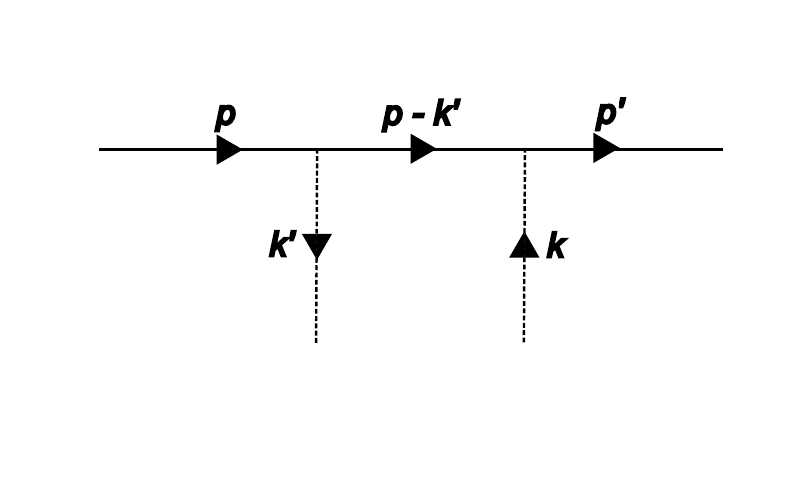}
\caption{$e^{-}\varphi\rightarrow e^{-}\varphi$}
\label{4-scattering-2}
\end{center}
  \end{minipage}
\end{figure}

First, we consider the 4-point amplitude $e^{-}\varphi\rightarrow e^{-}\varphi$. If we define $\varphi_{i}\,(i=1,2,3,4)$ as follows, 
\begin {align}
        \phi_{1} &= \int d^{4}x_{1}\,e^{-ik'x_{1}}\,c(x_{1})\,, \\ 
        \phi_{2} &= \int d^{4}x_{2}\,\overline{\theta}_{\alpha}(x_{2})\,\xi\,u_{s}({\bf p})_{\alpha}\,e^{ipx_{2}}\,,\\
        \phi_{3} &= \int d^{4}x_{3}\,\overline{u}_{s'}({\bf p}')_{\beta}\,e^{-ip'x_{3}}\,\overline{\eta}\,\theta_{\beta}\,(x_{3})\,,\\
        \phi_{4} &= \int d^{4}x_{4}\,e^{ikx_{4}}\,c(x_{4})\,,
\end {align}
where $\xi$ and $\overline{\eta}$ are degree-odd parameters, then we obtain
\begin{equation}
	\mathcal{A}_{4} = (2\pi)^{4}\,i\,g^{2}\,\delta^{4}(k+p-k'-p')\,\overline{u}_{s'}({\bf p}')\,[\tilde{S}(p-k')+\tilde{S}(k+p)]u_{s}({\bf p})\,\overline{\eta}\,\xi\,.
\end{equation}
This completely agrees with the result using the Feynman diagram (Figure~\ref{4-scattering-1},~\ref{4-scattering-2}).

\begin{figure}[H]
  \begin{minipage}[b]{0.45\linewidth}
 \begin{center}     
\includegraphics[width =7.5 cm]{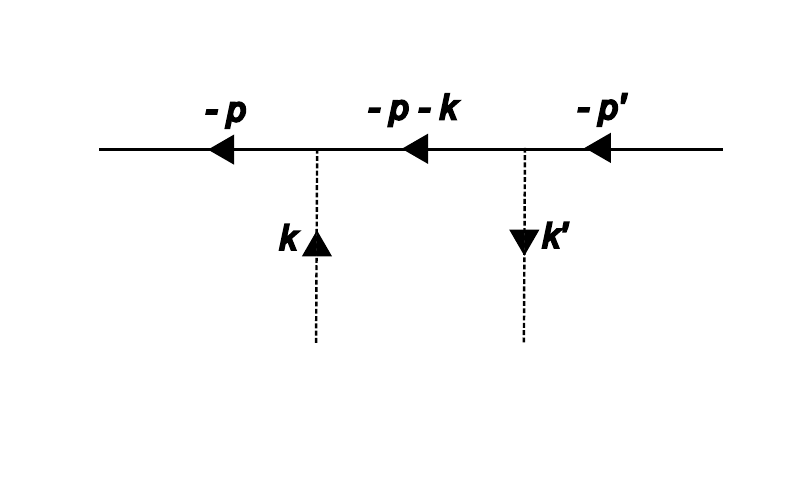}
\caption{$e^{+}\varphi\rightarrow e^{-}\varphi$}
\label{4-scattering-3}
\end{center}
  \end{minipage}
  \begin{minipage}[b]{0.45\linewidth}
   \begin{center}      
\includegraphics[width =7.5 cm]{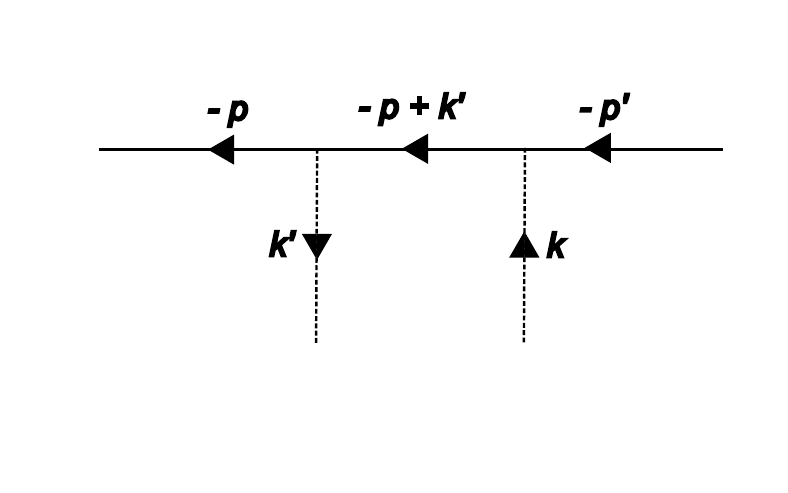}
\caption{$e^{+}\varphi\rightarrow e^{-}\varphi$}
\label{4-scattering-4}
\end{center}
  \end{minipage}
\end{figure}

Second, we consider the 4-point amplitude $e^{+}\varphi\rightarrow e^{+}\varphi$. If we define $\varphi_{i}\,(i=1,2,3,4)$ as follows, 
\begin {align}
        \phi_{1} &= \int d^{4}x_{1}\,e^{-ik'x_{1}}\,c(x_{1})\,, \\ 
        \phi_{2} &= {}-\int d^{4}x_{2}\,e^{ipx_{2}}\,\overline{v}_{s}({\bf p})_{\alpha}\,\overline{\eta}\,\theta_{\alpha}(x_{2})\,,\\
        \phi_{3} &= {}-\int d^{4}x_{3}\,\overline{\theta}_{\beta}(x_{3})\,\xi\,v_{s'}({\bf p}')_{\beta}\,e^{-ip'x_{3}}\,,\\
        \phi_{4} &= \int d^{4}x_{4}\,e^{ikx_{4}}\,c(x_{4})\,,
\end {align}
where $\xi$ and $\overline{\eta}$ are degree-odd parameters, then we obtain
\begin{equation}
	\mathcal{A}_{4} = -(2\pi)^{4}\,i\,g^{2}\,\delta^{4}(k+p-k'-p')\,\overline{v}_{s}({\bf p})\,[\tilde{S}(-p-k)+\tilde{S}(k'-p)]v_{s'}({\bf p}')\,\xi\,\overline{\eta}\,.
\end{equation}
This completely agrees with the result using the Feynman diagram (Figure \ref{4-scattering-3}, \ref{4-scattering-4}).

\begin{figure}[htbp]
  \begin{minipage}[b]{0.45\linewidth}
 \begin{center}     
\includegraphics[width =7.5 cm]{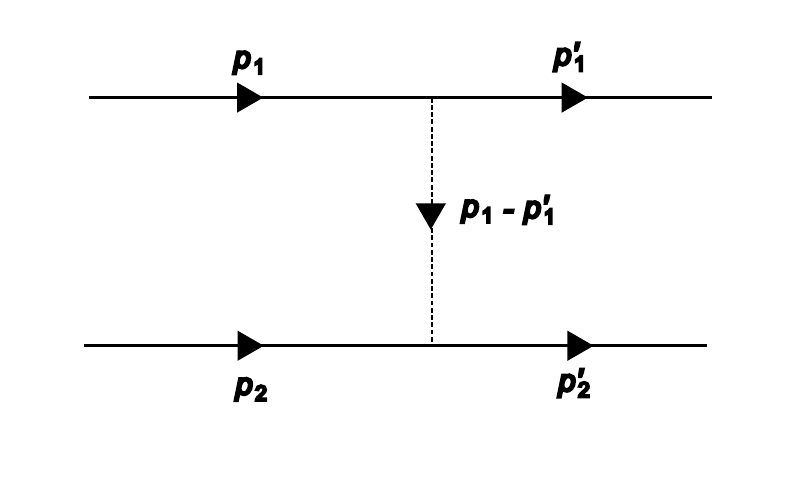}
\caption{$e^{-}e^{-}\rightarrow e^{-}e^{-}$}
\label{4-scattering-5}
\end{center}
  \end{minipage}
  \begin{minipage}[b]{0.45\linewidth}
   \begin{center}      
\includegraphics[width =7.5 cm]{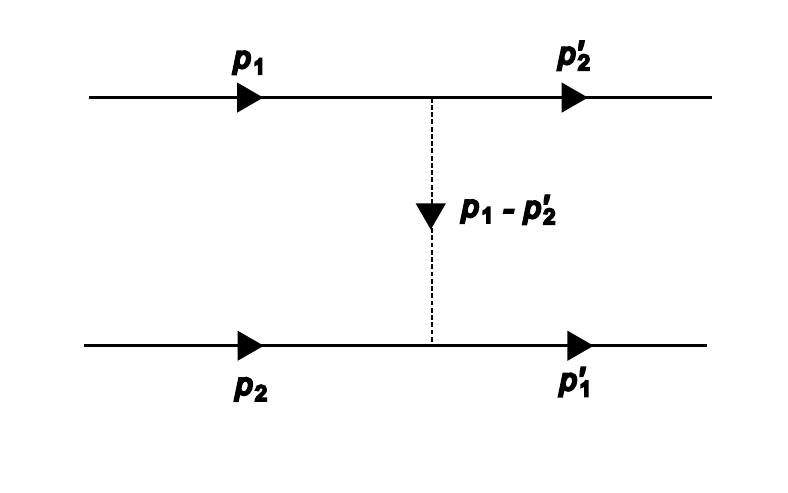}
\caption{$e^{-}e^{-}\rightarrow e^{-}e^{-}$}
\label{4-scattering-6}
\end{center}
  \end{minipage}
\end{figure}

Third, we consider the 4-point amplitude $e^{-}e^{-}\rightarrow e^{-}e^{-}$. If we define $\varphi_{i}\,(i=1,2,3,4)$ as follows, 
\begin {align}
        \phi_{1} &= \int d^{4}x_{1}'\,\overline{u}_{s_{1}'}({\bf p}_{1}')_{\alpha'}\,e^{-ip_{1}'x_{1}'}\,\overline{\eta}_{1}\,\theta_{\alpha'}(x'_{1}) \,,\\ 
        \phi_{2} &= \int d^{4}x_{2}'\,\overline{u}_{s_{2}'}({\bf p}_{2}')_{\beta'}\,e^{-ip_{2}'x_{2}'}\,\overline{\eta}_{2}\,\theta_{\beta'}(x'_{2})\,,\\
        \phi_{3} &= \int d^{4}x_{1}\,\overline{\theta}_{\alpha}(x_{1})\,\xi_{1}\,u_{s_{1}}({\bf p_{1}})_{\alpha}\,e^{ip_{1}x_{1}}\,,\\
        \phi_{4} &= \int d^{4}x_{2}\,\overline{\theta}_{\beta}(x_{2})\,\xi_{2}\,u_{s_{2}}({\bf p_{2}})_{\beta}\,e^{ip_{2}x_{2}}\,,
\end {align}
where $\overline{\eta}_{1}, \overline{\eta}_{2}, \xi_{1}$ and $\xi_{2}$ are degree-odd parameters, then we obtain
\begin{equation}
	\mathcal{A}_{4} = (2\pi)^{4}\,i\,g^{2}\,\delta^{4}(p_{1}+p_{2}-p_{1}'-p_{2}')\,\left[\frac{(\overline{u'}_{1}u_{1})(\overline{u'}_{2}u_{2})}{(p_{1}-p_{1}')^{2}+M^{2}}-\frac{(\overline{u'}_{2}u_{1})(\overline{u'}_{1}u_{2})}{(p_{1}-p_{2}')^{2}+M^{2}}\right]\,\overline{\eta}_{2}\,\overline{\eta}_{1}\,\xi_{1}\,\xi_{2}\,.
\end{equation}
This completely agrees with the result using the Feynman diagram (Figure \ref{4-scattering-5}, \ref{4-scattering-6}).

\begin{figure}[htbp]
  \begin{minipage}[b]{0.45\linewidth}
 \begin{center}      
\includegraphics[width =7.5 cm]{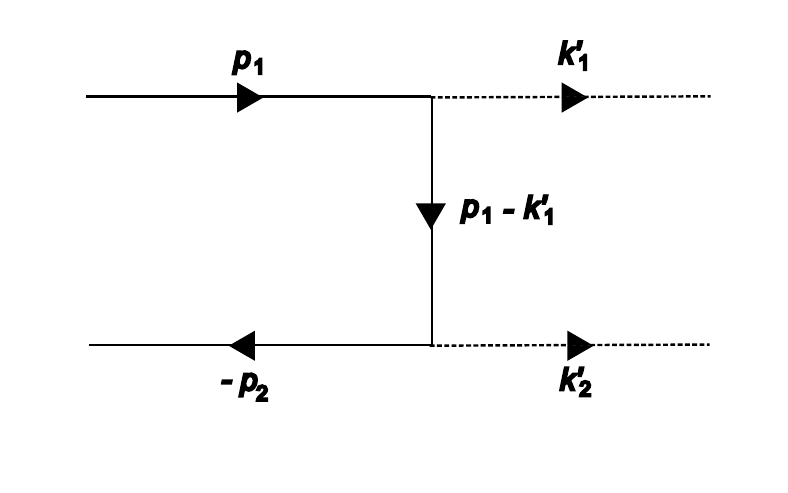}
\caption{$e^{+}e^{-}\rightarrow \varphi\varphi$}
\label{4-scattering-7}
\end{center}
  \end{minipage}
  \begin{minipage}[b]{0.45\linewidth}
   \begin{center}      
\includegraphics[width =7.5 cm]{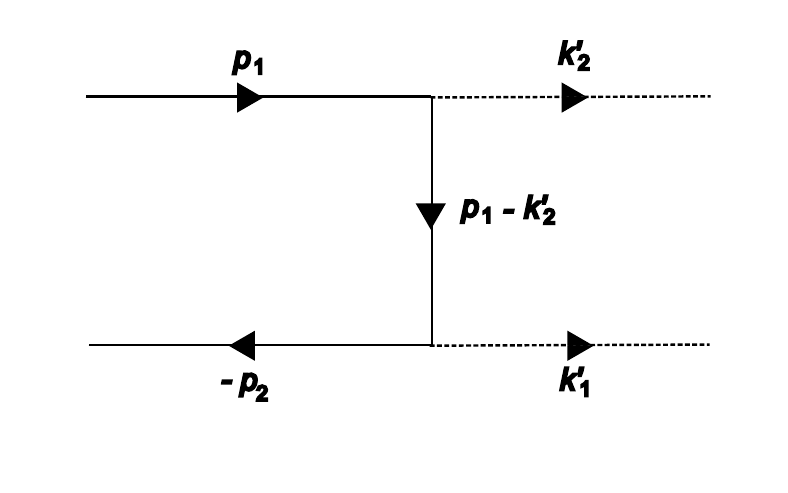}
\caption{$e^{+}e^{-}\rightarrow \varphi\varphi$}
\label{4-scattering-8}
\end{center}
  \end{minipage}
\end{figure}

Fourth, we consider the 4-point amplitude $e^{+}e^{-}\rightarrow \varphi\varphi$. If we define $\varphi_{i}\,(i=1,2,3,4)$ as follows, 
\begin {align}
        \phi_{1} &=  \int d^{4}x_{1}\,e^{-ik_{1}'x_{1}}\,c(x_{1})\,, \\ 
        \phi_{2} &=  \int d^{4}x_{2}\,e^{-ik_{2}'x_{2}}\,c(x_{2})\,, \\ 
        \phi_{3} &={}-\int d^{4}x_{3}\,e^{ip_{1}x_{3}}\,\overline{v}_{s_{1}}({\bf p}_{1})_{\alpha}\,\overline{\eta}\,\theta_{\alpha}(x_{3})\,,\\
        \phi_{4} &= \int d^{4}x_{4}\,\overline{\theta}_{\beta}(x_{4})\,\xi\,u_{s_{2}}({\bf p}_{2})_{\beta}\,e^{ip_{2}x_{4}}\,.
\end {align}
where $\xi$ and $\overline{\eta}$ are degree-odd parameters, then we obtain
\begin{equation}
	\mathcal{A}_{4} = {}-(2\pi)^{4}\,i\,g^{2}\,\delta^{4}(p_{1}+p_{2}-k_{1}'-k_{2}')\,\overline{v}_{s_{2}}({\bf p}_{2})[\tilde{S}(-k_{1}'+p_{1})+\tilde{S}(-k_{2}'+p_{1})]\,u_{s_{1}}({\bf p}_{1})\,\overline{\eta}\,\xi\,.
\end{equation}
This completely agrees with the result using the Feynman diagram (Figure \ref{4-scattering-7}, \ref{4-scattering-8}).

\begin{figure}[htbp]
  \begin{minipage}[b]{0.45\linewidth}
 \begin{center}      
\includegraphics[width =7.5 cm]{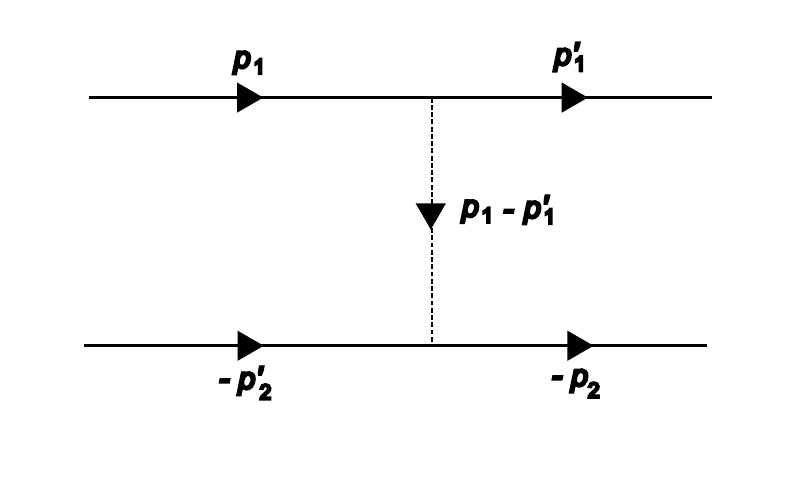}
\caption{$e^{+}e^{-}\rightarrow e^{+}e^{-}$}
\label{4-scattering-9}
\end{center}
  \end{minipage}
  \begin{minipage}[b]{0.45\linewidth}
   \begin{center}      
\includegraphics[width =7.5 cm]{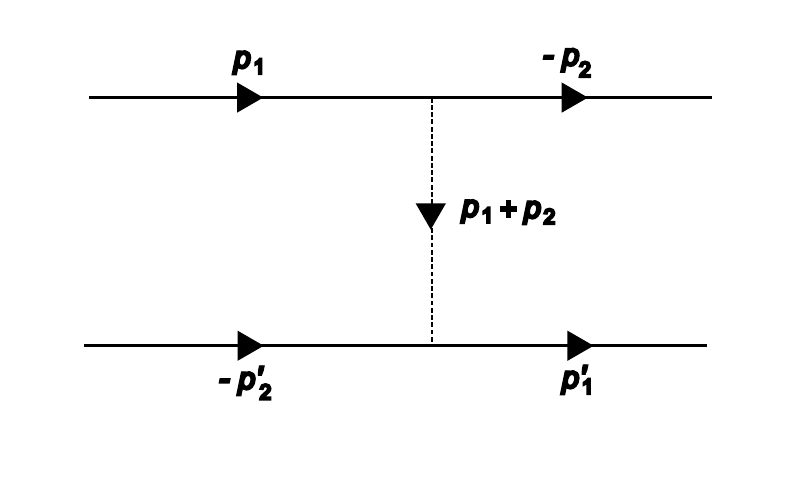}
\caption{$e^{+}e^{-}\rightarrow e^{+}e^{-}$}
\label{4-scattering-10}
\end{center}
  \end{minipage}
\end{figure}

Finally, we consider the 4-point amplitude $e^{+}e^{-}\rightarrow e^{+}e^{-}$. If we define $\varphi_{i}\,(i=1,2,3,4)$ as follows, 
\begin {align}
        \phi_{1} &= \int d^{4}x_{1}'\,\overline{u}_{s_{1}'}({\bf p}_{1}')_{\alpha'}\,e^{-ip_{1}'x_{1}'}\,\overline{\eta}_{1}\,\theta_{\alpha'}(x'_{1}) \,,\\ 
        \phi_{2} &=  {}-\int d^{4}x_{2}'\,\overline{\theta}_{\beta'}(x_{2}')\,\xi_{2}\,v_{s_{2}'}({\bf p}_{2}')_{\beta'}\,e^{-ip_{2}'x_{2}'}\,,\\
        \phi_{3} &=  \int d^{4}x_{3}\,\overline{\theta}_{\alpha}(x_{3})\,\xi_{1}\,u_{s_{1}}({\bf p}_{1})_{\alpha}\,e^{ip_{1}x_{3}}\,,\\
        \phi_{4} &= {}-\int d^{4}x_{4}\,e^{ip_{2}x_{4}}\,\overline{v}_{s_{2}}({\bf p}_{2})_{\beta}\,\overline{\eta}_{2}\,\theta_{\beta}(x_{4})\,.
\end {align}
where $\overline{\eta}_{1}, \overline{\eta}_{2}, \xi_{1}$ and $\xi_{2}$ are degree-odd parameters, then we obtain
\begin{equation}
	\mathcal{A}_{4} = {}-(2\pi)^{4}\,i\,g^{2}\,\delta^{4}(p_{1}+p_{2}-p_{1}'-p_{2}')\,\left[\frac{(\overline{u}_{1}'v_{2}')(\overline{v}_{2}u_{1})}{(p_{1}+p_{2})^{2}+M^{2}}-\frac{(\overline{u}_{1}'u_{1})(\overline{v}_{2}'v_{2})}{(p_{1}-p_{1}')^{2}+M^{2}}\right]\,\xi_{2}\,\overline{\eta}_{1}\,\overline{\eta}_{2}\,\xi_{1}\,.
\end{equation}
This completely agrees with the result using the Feynman diagram (Figure \ref{4-scattering-9}, \ref{4-scattering-10}).

\break

\end{document}